\newcommand{\ignore}[2]{\hspace{0in}#2}

\documentclass[acmsmall]{acmart}

\def\BibTeX{{\rm B\kern-.05em{\sc i\kern-.025em b}\kern-.08emT\kern-.1667em\lower.7ex\hbox{E}\kern-.125emX}}
    
\begin{document}

\settopmatter{printacmref=false}
\setcopyright{none}
\renewcommand\footnotetextcopyrightpermission[1]{}
\pagestyle{plain}

%

\title{Multi-Criteria-based Dynamic User Behaviour Aware Resource Allocation in Fog Computing}
%
\author{Ranesh Kumar Naha}
\authornotemark[1]
\email{raneshkumar.naha@utas.edu.au}
\orcid{0000-0003-4165-9349}
\author{Saurabh Garg}
\authornotemark[1]
\email{saurabh.garg@utas.edu.au}
\affiliation{%
  \institution{University of Tasmania}
  \streetaddress{School of Technology, Environments and Design}
  \city{Hobart}
  \state{Tasmania}
  \postcode{7001}
}



 




%

%
\begin{abstract}
Fog computing is a promising computing paradigm in which IoT data can be processed near the edge to support time-sensitive applications. However, the availability of the resources in the computation device is not stable since they may not be exclusively dedicated to the processing in the Fog environment. This, combined with dynamic user behaviour, can affect the execution of applications. To address dynamic changes in user behaviour in a resource limited Fog device, this paper proposes a Multi-Criteria-based resource allocation policy with resource reservation in order to minimise overall delay, processing time and SLA violation which considers Fog computing-related characteristics, such as device heterogeneity, resource constraint and mobility, as well as dynamic changes in user requirements. We employ multiple objective functions to find appropriate resources for execution of time-sensitive tasks in the Fog environment. Experimental results show that our proposed policy performs better than the existing one, reducing the total delay by 51\%. The proposed algorithm also reduces processing time and SLA violation which is beneficial to run time-sensitive applications in the Fog environment.

\end{abstract}


%

%
\maketitle

\section{Introduction}
We are approaching a well-connected modern world in which each and every person and things will be connected to the internet. According to Statista, the total number of connected devices will be more than 75 billion by 2025  \cite{Statista}. By 2022, the monthly mobile internet traffic alone will be 77 exabytes, as predicted by Cisco VNI Global Mobile Data Forecast \cite{CiscoForecast}. The sudden rise of IoT devices and internet traffic will face unforeseen anarchy if we do not focus on data processing near to the users. This is the reason why the Fog computing concept has been developed. It is a distributed computing paradigm in which applications are processed near to users or near to the edge by using either edge devices or utilising the idle computation power of various autonomous devices \cite{naha2020deadline}. According to Shi et al. \cite{shi2016edge} edge computing and Fog computing are interchangeable with each other but edge is more towards things-centric while Fog is mostly infrastructure-centric. The Fog computing environment consists of undedicated servers for application processing, unlike previous computing paradigm such as the Cloud  \cite{battula2019micro}. Any devices, such as routers, switches, RoadSide Units (RSUs), smartphones, tablets, laptop and stationary computation devices that have computation power can be part of Fog computation and processing  \cite{naha2018fog,naha2018fogarc}.

Clearly such devices are heterogeneous, not dedicated devices and connected using multiple type of network connections \cite{garcia2018introducing}. Moreover, the user behaviour in the Fog environment can also be dynamic. Users might change their requirements even after submitting an application request based on their current situation. For example, real-time traffic application and augmented reality applications are the two examples of such applications in which users' behaviour might vary constantly, based on their changing requirements, as well as the current situation  \cite{naha2018fog}.       

\ignore{Importance is solving the problems}
The problems of dynamic user behaviour and the unstable characteristics of available resources in the devices make resource allocation a difficult task. Resource allocation in the Fog should identify and select resources, and consider these issues for efficient and reliable execution of the application. If the application is for an emergency system, then failing to complete the tasks on time will incur financial loss and may even cause a threat to human life (for example, the emergency fire response service, driverless cars, emergency vehicle management, and many more). 

Most research \cite{mahmud2018quality,aazam2016mefore,skarlat2017towards,lin2015cloud} on Fog computing neglected the combination of user and resource challenges as regards dynamic user behaviour. To fill this gap, this research aims to develop a resource allocation technique to minimise user and resource related challenges. In summary, this research has the following contributions:

\begin{enumerate}
    \item To propose a solution to the application placement problem while user behaviour is changing dynamically; 
    \item To propose Multi-Criteria (MC) based resource allocation for Fog applications which will consider multiple Fog characteristics, and 
    \item To introduce a previous task submission history-based resource reservation technique to deal with the time-sensitivity of applications.
 \end{enumerate}
The rest of the paper is organised as follows. Section 2 presents the literature about Fog resource allocation and scheduling. Section 3 describes the system scenario in smart transportation and augmented reality related applications. Next, Section 4 describes the problem and solution approach. Section 5 presents the details on the methodology by covering proposed algorithms, network models, pricing models and performance metrics. The simulation and the experimental setup are discussed in Section 6. Section 7 presents experimental results and discussion, and finally, Section 8 concludes this article and suggests future research direction.


\section{Related Work}

The resource allocation problem has been studied in various computing paradigms such as the Cloud and Fog computing environment. In this section, we present a state-of-the-art solution for the problem of resource allocation for time sensitive applications in the Fog and other distributed computing environments. Resource allocation policies used in other distributed computing paradigms do not fit exactly with Fog computing because of the dynamic nature of the users and resources in the Fog environment. 

A number of research studies about resource allocation in the Fog has been undertaken in the last few years Most of these have considered user requirements and network throughput for Fog resource allocation. Some work focuses on Cloud-Fog resource allocation while others only focus on Fog resource allocation. We summarise the current works and, at the end of this section, describe how our works would differ from others.

 
 
\subsection{Resource Allocation in Cloud and Distributed Computing}
Resource allocation policies are extensively studied in the Cloud computing paradigm because costs is the most important factor for the Cloud users. Ghanbari and Othman \cite{ghanbari2012priority} proposed a three level, priority-based scheduling algorithm for the Cloud. Their proposed algorithm considered the objective level, the attributes level and alternative level priority. However, priority in multiple characteristics of the resources is not considered in their work. 

Baranwal et al. \cite{baranwal2015fair} proposed double auction-based resource allocation in which they employed multi-attribute combinations without considering the cost. However, their work implemented auction fairness to ensure the satisfaction of the users and also considered provider reputation to avoid spurious QoS assurance. Xu et al. \cite{xu2013qos} proposed a Cloud resource allocation technique for cost- and time-sensitive users. Their proposed algorithm is based on cooperative game theory. However, future availability of the resources is not considered in their work. Zheng and Shroff \cite{zheng2016online} studied the online scheduling problem for deadline-sensitive jobs in the Cloud environment. Their work considered multiple resource sharing to meet application deadlines and proposed an online algorithm to deal with deadline sensitivity. 

Some research work specifically focused on user behaviour in the distributed computing environment. Schlagkamp et al. \cite{schlagkamp2016understanding} investigated the differences and similarities in user job submission behaviour in High Throughput Computing (HTC) and High-Performance Computing (HPC). Their findings show that modelling user-based HTC job submission behaviour requires knowledge of the underlying bags of tasks, which is often unavailable. Also suggested was in-depth characterisation of waiting times in order to improve the correlation analysis between queueing times and the subsequent user job submission behaviour. Panneerselvam et al.  \cite{panneerselvam2017inot} proposed a novel prediction model named InOt-RePCoN (Influential Outlier Restrained Prediction with Confidence Optimisation) which is aimed at a tri-fold forecast for predicting the expected number of job submissions, the session duration for users, and also the job submission interval for the incoming workloads. The proposed framework exploits the AutoregRessive Integrated Moving Average (ARIMA) technique, integrated with a confidence optimiser for prediction. It achieves a reliable level of accuracy in predicting user behaviours by way of exploiting the inherent periodicity and predictability of every individual job of every single user. It can predict user behaviour trends but it fails to address the problem of dynamic changes in user behaviour.

\subsection{Resource Allocation in Fog Computing}
Bittencourt et al. \cite{bittencourt2017mobility} proposed mobility-aware application scheduling by considering the hierarchical composition of the Fog and Cloud environments. However, they did not consider other unique characteristics of the Fog computing environment, such as CPU availability fluctuation and future arrival of the applications. 
Silva and Fonseca \cite{da2018resource} proposed Fog-Cloud resource allocation which would consider available system resources. Their proposed mechanism is known as Gaussian Process Regression for Fog-Cloud Allocation (GPRFCA). GPRFCA analyses previous submission requests and latency for resource allocation. It also minimises the energy consumption of Fog nodes. In another work, Du et al. \cite{du2018computation} proposed resource allocation for the Fog-Cloud environment by ensuring min-max fairness. The main goal of that work is to optimise transmitted power, radio bandwidth and computation resources by guaranteeing maximum tolerable delay and user fairness. Their proposed resource allocation algorithm is known as Computation Offloading and Resource Allocation algorithm (CORA).

Throughput and load balancing are also important for resource allocation in the Fog. Yu et al. \cite{yu2018green} proposed a resource allocation technique by employing three different algorithms (for example, Modified Distributed Inner Convex Approximation, Joint Benders Decomposition and Dinkelbach Algorithm) for Fog resources. The objective of that is to maximise the utility function by considering network throughput. Another work by Yu et al. \cite{yu2018green1} 
introduces the alternating direction method of multipliers (ADMM) and Branch-and-Bound algorithms, along with Joint Benders Decomposition and Dinkelbach Algorithm for large-scale resource allocation in the Fog environment. For this large-scale allocation, throughput was considered, based on the distance of Fog nodes. Xu et al. \cite{xu2018dynamic} proposed using the Dynamic Resource Allocation Method (DRAM) to balance the load in the Fog environment. In this method, they considered resource requirements and spare space in computation nodes, in order to determine the dynamic resource scheduling and static resource allocation.

Zhang and Li \cite{zhang2018enabling} proposed a resource allocation technique that preserves the privacy of Fog nodes from unwanted attacks, even if the private key of all Fog nodes is corrupted. In their proposed scheme, a group public key will be generated by a group of Fog devices that will make the system secure against eavesdroppers and that will be used as a smart gateway.  Garc\'ia-Valls et al. \cite{garcia2017adjusting} discussed time-sensitivity of the cyber physical system due to dynamic behaviour enforcement. Along with other techniques, they suggested setting priority to the servers. However, it is required to analyse extensively the prioritisation of Fog devices since the devices in the Fog environment are more dynamic. In another work,  Garc\'ia-Valls et al. \cite{garcia2018accelerating} proposed an execution model using Fog servers for smart eHealth services. The key idea of this work is offloading processes to the multicore Fog servers to accelerate the processing. However, the main focus of our work is the consideration of user behaviour and Fog characteristics.

\begin{table*}[ht]
	\centering
    \footnotesize
	\caption{Comparison of various resource allocation method in Fog computing environment}
	\label{tablrcom}
	\begin{tabular}{L{1.6cm}C{1.4cm}C{1.4cm}C{1.2cm}C{1.4cm}C{1.4cm}C{1.4cm}C{1.2cm}} 
		\toprule
Research Work	& Device heterogeneity	& Resource constraint & Device mobility & CPU availability fluctuation & Completion time & Application time-sensitiveness & Future arrival \\ \hline
\hline
Aazam et al. (2016) \cite{aazam2016mefore}  &	\checkmark	&	\checkmark	&		&		&	\checkmark	&	&	\\ \hline
Skarlat et al. \cite{skarlat2017towards} (2017) &	\checkmark	&	\checkmark	&		&		&	\checkmark	&	\checkmark & 	\\ \hline
Bittencourt et al. \cite{bittencourt2017mobility} (2017) &	\checkmark	&	\checkmark	&	\checkmark	&		&		&	 & 	\\ \hline
Silva and Fonseca (2018) \cite{da2018resource}	&	\checkmark	&	\checkmark	&		&		&	\checkmark	& 	& \checkmark	\\ \hline
Du et al. (2018) \cite{du2018computation}	&	\checkmark	&	\checkmark	&		&		&		&	&	\\ \hline
Yu et al. (2018) \cite{yu2018green}	&	\checkmark	&	\checkmark	&		&		&		&		\\ \hline
Xu et al. (2018) \cite{xu2018dynamic}	&	\checkmark	&	\checkmark	&		&		& \checkmark		&	&	\\ \hline
Zhang and Li (2018) \cite{zhang2018enabling}	&	\checkmark	&	\checkmark	&		&		&		&	&  \\ \hline
Garc\'ia-Valls et al. (2018) \cite{garcia2018accelerating} &	\checkmark	&	\checkmark	&		&		&	\checkmark	&	\checkmark &	\\ \hline
Mahmud et al. \cite{mahmud2018quality} (2018) &	\checkmark	&	\checkmark	&		&		&	\checkmark	&	\checkmark &	\\ \hline
MC-Based (This work) &	\checkmark	&	\checkmark	&	\checkmark	&	\checkmark	&	\checkmark	&	\checkmark & \checkmark	\\ \hline

\bottomrule
\end{tabular}
\end{table*}

Mahmud et al. \cite{mahmud2018quality} proposed a Quality of Experience (QoE) aware application placement policy which prioritises different application placement by using fuzzy logic. In this, they considered service access and round-trip time, resource requirements and availability, and processing time and speed, while placing an application into the Fog infrastructure. Using fuzzy logic, the Rating of Expectation (RoE) and Capacity Class Score (CCS) have been calculated before placing an application in the Fog infrastructure. QoE aware application placement is compared with MeFoRE \cite{aazam2016mefore}, QoS-aware \cite{skarlat2017towards} and CloudFog \cite{lin2015cloud} which showed that the proposed QoE-aware policy would perform better on processing time, service quality, infrastructure cost and network traffic. Hence, we are going to compare our proposed MC-based resource allocation with QoE-aware policy as a validation of our proposed policy. Comparison of various resource allocation methods is presented in Table \ref{tablrcom}.

From the above literature, it is clear that no one has considered the unique characteristics of the Fog such as CPU availability fluctuation, dynamic behaviour of users and the mobility of the available Fog resources. In this research, we considered the above constraints, along with access time, requirements of resources, and processing time in our proposed policy to solve the dynamic user behaviour problem in the Fog environment. These constraints will help in the successful completion of application requests in the highly unstable available resources for Fog application processing in a time-sensitive manner.

\ignore{+++++=====Need to rewrite form here=====+++++}


%



 
\section{System Scenario}

If IoT based smart transportation and augmented reality related applications depend on the Cloud service, then time sensitivity of the applications may not be met. Usually, the communication latency of the Cloud service is high, and also real-time interaction is hard  \cite{mahmud2018cloud}. It is also assumed that the network latency in the Cloud is 100 milliseconds (ms) from the source \cite{mahmud2018cloud}. However, users of such applications would constantly need real-time interactions. Processing requests near to the users is a good solution that can be undertaken by the Fog computing environment. Resources in the Fog are dynamic with respect to utilisation, location and availability. Fog middleware has to handle the dynamic nature of the resources, as well as the dynamic nature of the users. Further descriptions of smart transportation and augmented reality related applications are presented below. However, the proposed solution that is presented in this paper would cope with all types of time-sensitive applications in the Fog. 

\subsection{Special Occurrence in Smart Transportation System}
For the IoT based transportation system, the application is required to deal with some frequent decisions, for example, traffic avoidance, action on roadblocks or accidents, and making way for emergency vehicles. These types of decisions might be useless if the delay is even a few seconds. In smart transportation application, each vehicle is considered a user. Assuming that our application has the functionality to declare a vehicle as an emergency vehicle and based on the risk parameters, the level of emergency can be varied by the users. If the scenario is some emergency situation caused by fire, every second is very important. A smart firefighting team would be able to deal with such emergency situations more efficiently. In such cases, the deadline is an important factor according to the changing behaviour of the users. Based on the action of smart firefighting, the amount of damage can be minimised.

One of the other examples is a smart traffic application which is based on the changing behaviour of users for congestion avoidance and fuel cost minimisation. By using the information from the connected vehicle and other roadside sensors, it is possible to minimise fuel cost and congestion. In a smart connected transportation system, every vehicle is equipped with a dash cam. By processing the data obtained from the dash cam and roadside cameras, we should be able to achieve various goals. The dash cam could supply information about the current location of the vehicle, road conditions, traffic conditions, special occurrences (roadblocks or accidents), weather conditions at current location, and current speed. Other application-related metadata, such as traffic updates, vehicle fuel consumption characteristics, environmental factors (temperature, light, urban or rural area, peak or off-peak traffic status) will be retrieved from the local Fog server to process the application output. Also, some input will be generated from other roadside units (RSU). Fog devices will be able to process all of the data to generate the output which basically indicates the driving path, estimated fuel consumption or any other emergency occurrence. The input behaviour of the users and other sources is dynamic. Due to this, the output is always transformed dynamically, over time. Users might place a sudden request to minimise the cost of the Fog service but not need the congestion avoidance and fuel cost minimisation services. To deal with such situations, the system should find a way to minimise the cost of using Fog applications. 

\subsection{Augmented Reality Related Applications}
In augmented reality, users have interactive interaction with the real-world objects with the most dynamic behaviour. User experience is important in these kinds of applications. Fog computing can be used to enable better user experience with augmented reality by processing application requests near the users. An augmented reality application for tourists should provide a seamless experience. Tourists, generally, love to explore historical or iconic places. They would also want to know about the history behind the different landmarks and architecture. In such a scenario, if the processing is done in the Cloud, it may not provide a seamless experience. Similarly, the user experience will not be good in augmented reality-based on-line games using the Cloud. For any augmented reality application, user behaviour will always change dynamically. In the case of augmented reality application for tourists, the user behaviour is always changing, based on the amount of information they request and the level of detail sought with each request. Augmented reality-based multi-player online game user requirements can also change dynamically, based on the level that they are playing and the complexity of that level.

In both scenarios stated above, it is not possible to predict user behaviour while the requirements of the users are changing over time. We need to understand the pattern of user behaviour so that we can manage an application for the users to have a seamless service experience. We need to identify how frequent user behaviour is changing and in which situations it is changing, so that applications can handle this dynamism.


It is obvious that Fog devices will do the processing for the application and store some intermediate information. Once the application processing is done, the Fog device may save some information to the Fog server which would be necessary for the next few hours. For long-term storage, processed information will be stored in the Cloud for the future. Multiple application requests could be sent by an end device to Fog devices for processing. These applications will be scheduled by various Fog devices in order to achieve minimum latency.

\subsection{Resource Model}

The processing devices in the Fog environment are Fog devices and Fog servers. These devices contain processing power, RAM and network connectivity. More details about the resources are presented in Table \ref{tsFogSimPara}. The device that has more computational resources compared with Fog devices is designated as the Fog server. These devices support virtualisation with Virtual machines. Multiple Fog devices form a Fog cluster and every Fog cluster is connected to a Fog server. Fog devices and servers are connected to the Cloud, in order store data for long term processing. Fog devices are responsible for IoT application processing which can also be defined as Fog processing. Each IoT application consists of multiple tasks; more details about tasks are presented in Table \ref{tsFogSimPara}.

\subsection{Application scheduling on Fog Infrastructure}
\begin{figure}[!t]
	\centering
	\includegraphics[width=3in]{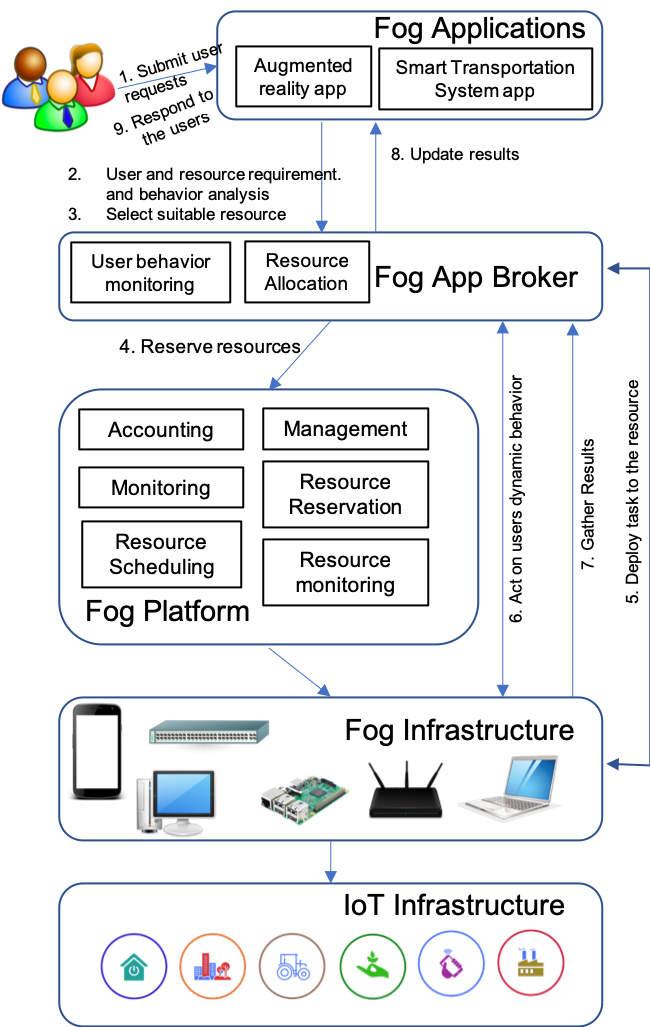}
	\caption{Application scheduling in Fog environment.}
	\label{SysSc}
\end{figure}

In order to satisfy time-sensitivity requirements, applications need to be executed in the Fog infrastructure. Fig. \ref{SysSc} shows the application scheduling and execution, using Fog resources in the Fog computing environment. Any computation device which is willing to share its resources for computation can be registered in the Fog environment and the owner of the device will receive incentives based on its contribution \cite{battula2019micro}. On the other hand, if the device is unable to serve the request after the agreement, then the device owner should pay the penalty. In our application scenario, users submit the request to the application for further processing. The Fog app broker evaluates selected resources for allocation by analysing the requirements and behaviour of the users and the resources. 

An application broker is a mediator between the Fog application and the Fog platform. It selects appropriate resources with the help of the Fog platform. The Fog app broker also reserves resources based on users' behaviour. The Fog platform caters for resource scheduling, monitoring, reservation and management. The application is deployed in the Fog infrastructure. The Fog platform helps the Fog app broker to select appropriate resources for resource allocation. This platform monitors available resources in the Fog environment. After selecting resources, the Fog platform schedules application to the Fog resources and continues to monitor them. If any dynamic changes are initiated by the user, the Fog platform reschedules the application to the Fog resources, based on the changes in users' requirements. The management of application and Fog infrastructure is taken care of by the Fog platform. It also reserves resources based on application submission history in the Fog environment. Moreover, all accounting related operations, such as billing and usage, are maintained by the Fog platform. A detailed description about resources allocation, reservation and user dynamic behaviour handling is presented in section \ref{pro_pol}. The final outcome of the application is collected from the infrastructure and passed to the application, in 
response to the application for the submitted request.
  
\begin{table*}[htbp]
\centering
\tiny
\caption{Notations}
\label{tabnota}
\begin{tabular}{L{1.5cm}L{4.3cm}L{1.5cm}L{4.3cm}} 
\toprule
Symbol & Definition & Symbol & Definition \\ \hline
$\alpha$ & Constant value for penalty & $\beta$ & Penalty rate \\ \hline
$A_b$ & Available battery charge &
$A_{dr}$ & Battery discharging rate by the application \\ \hline
$A_n$ & Constant for processing and transmission delay &
$A_s$ & Availability score \\ \hline
$AT_{cost}$ & Total cost per application &
$A_v$ & Availability based on battery power \\ \hline
$AB_{WL}$ & Available bandwidth of a link &
$AB_{WP}$ & Available bandwidth of a path \\ \hline
$b_w$ & Bandwidth &
$B_{WL}$ & Bandwidth of a link \\ \hline
$B_{WP}$ & Bandwidth of a path &
$C$ & Link capacity \\ \hline
$C_A, C_B, C_N$ & Connection A, B to N &
$C_{afs}$ & CPU availability fluctuation score \\ \hline
$C_c$ & Connection time to the Cloud &
$C_{FD}$ & Connection time to the Fog device \\ \hline
$C_{FS}$ & Connection time to the Fog server &
$CP$ & Unit price for connectivity \\ \hline
$CPU_{fr}$ & average CPU fluctuation rate &
$CPU_s$ & CPU speed in MIPS \\ \hline
$C_t$ & Completion time &
$CTA$ & Total completion time for an application instance \\ \hline
$CTA_{avg}$ & Average completion time of the system for each request &
$C_{total}$ & Total charges for connectivity \\ \hline
$CTU_{avg}$ & Average completion time of a single user request &
$D_s$ & Data size \\ \hline
$D^{fixed}$ & Minimum fixed delay &
$DIP_{avg}$ & Average internal communication delay \\ \hline
$DIP_{total}$ & Total internal communication delay &
$DP_{avg}$ & Average delay \\ \hline
$DP_{total}$ & Total delay &
$DT$ & Delay time \\ \hline
$E_t$ & Execution time &
$FD_x$ & Cost variable of Fog device \\ \hline
$FS_x$ & Cost variable of Fog server &
$Fr_r$ & fluctuation rate in each time-frame \\ \hline
$F_{rs}$ & Free resource score &
$G_d$ & Geographical distance \\ \hline
$h$ & Number of hops &
$J_s$ & Job size in MIPS \\ \hline
$k$ & Total number of intermediate links &
$L$ & Length of the frame \\ \hline
$M_c$ & Messaging request to the Cloud  &
$M_{FD}$ & Messaging request to the Fog device  \\ \hline
$M_{FS}$ & Messaging request to the Fog server  &
$M_t$ & Migration time \\ \hline
$M_{th}$ & Throughput of the medium &
$M_{total}$ & Total charges for messaging \\ \hline
$MP$ & Messaging unit price  &
$N_{dL}$ & Network delay of a link \\ \hline
$N_{dP}$ & Network delay of a path &
$N_u$ & Number of users sharing a link \\ \hline
$P_c$ & Processing request to the Cloud  &
$PC_{u}$ & Number of packets need to be sent to the Cloud \\ \hline
$P_d, \delta$ & Propagation delay &
$P_{d_n}$ & Propagation delay for a node \\ \hline
$P_{dp}$ & Propagation delay of a path &
$P_{FD}$ & Processing request to the Fog device  \\ \hline
$P_{FS}$ & Processing request to the Fog server  &
$P_{ip}$ & Total number of communications \\ \hline
$PR_d$ & Processing delay  &
$PR_{d_n}$ & Processing delay for a node  \\ \hline
$PR_dp$ & Processing delay of a path  &
$PP$ & Processing unit price  \\ \hline

$P_{total}$ & Total charges for rules engine  &
$P_u$ & Total number of communications sent by the user \\ \hline
$Q_d$ & Queuing delay  &
$Q_{d_n}$ & Queuing delay for a node  \\ \hline
$Q_{dp}$ & Queuing delay of a path &

$R_t$ & Response time \\ \hline
$r$ & Response of the request &
$SD_{max}$ & Maximum supported distance \\ \hline
$SP$ & Registry or Shadow unit price &
$SR_c$ & Registry or Shadow request to the Cloud in KB \\ \hline
$SR_{FD}$ & Registry or Shadow request to the Fog device in KB &
$SR_{FS}$ & Registry or Shadow request to the Fog server in KB  \\ \hline
$S_{total}$ & Total charges for shadow and registry &
$T_{bd}$ & Throughput based on distance \\ \hline
$TC$ & Total cost for an application instance with $n$ number of requests &
$TC_{req}$ & cost for using Cloud and Fog environments for a single request \\ \hline
$T_d$ & Transmission delay  &
$T_{d_n}$ & Transmission delay for a node  \\ \hline
$T_dp$ & Transmission delay of a path &
$TIP$ & total number of request and response \\ \hline
$TP$ & Number of packets transmitted &
$tP_{c}$ & Processing time required by Cloud \\ \hline
$tP_{fd}$ & Processing time required by Fog device &
$tP_{fs}$ & Processing time required by fog server \\ \hline
$tP_u$ & Total time required to send $P_u$ packet &
$tPC_u$ & Total time required to send $PC_u$ packet \\ \hline
$TPT$ & total processing time &
$T_r$ & Transmission rate \\ \hline
$t_h$  & Throughput &
$t_i, t_j, t_n$ & $t$ is Task, $i$ is $i^{th}$ task, $j$ is completed task, $n$ is the remaining task  \\ \hline 
$T_d$, $T_{t_n}$ & Task should be completed in $d$ time, $t_n$ time needed to complete the task &
$U$ & Data size unit \\ \hline

$W$ & Packet size \\ \hline
\bottomrule                               
\end{tabular}
\end{table*}







\section{Proposed MC-Based Resource Allocation}

Special occurrences in the smart transportation system and augmented reality related applications have been chosen as the use case. To solve the resource allocation problem for time-sensitive applications with deadline requirements and dynamic users' behaviour in the Fog environment, we need to find appropriate resources to finish the tasks that need to be completed by the application. We are going to propose a solution by suggesting an MC-Based resource allocation technique. The mathematical formulation is carried out to calculate delay and cost for the experiments and simulations. Finally, three performance metrics are described to evaluate the system performance. 

\subsection{Problem Description}
\label{ProDef}
This research is dealing with the following problem: How can deadlines be satisfied, based on user requirements for time-sensitive applications, considering the dynamic behaviour of users in the Fog computing environment? Satisfying these requirements is challenging because of the limited, heterogeneous and dynamic nature of the resources in the Fog computing environment compared with the Cloud. In the Fog, we have a limited number of heterogeneous devices. On top of that, there is no dedicated device which is responsible for Fog processing only. Most of the devices in the IoT environment are connected via wireless networks in which throughput is less compared with the wired network. Not only that, we do not have full control of the resources because we are using end devices and intermediate network devices for Fog processing. As handheld mobile devices are part of the Fog environment, they are moving from one network to another. 


Dynamic user behaviour-aware resource allocation in unstable Fog devices is a challenging task and we need to solve this problem to execute user application in the Fog environment. To address this problem, this research is going to investigate deadlines as the dynamic behaviour of users' requirements. Here, a deadline is referring to both scheduling time and the running time of the tasks submitted by the users. More precisely, the deadline is the time in which users would want to complete their tasks. To solve the above problem, efficient resource allocation and resources reservation are required. We need to have some intelligence to schedule time-sensitive applications in the Fog environment; otherwise, we will not be able to satisfy the time-sensitivity of the applications.



During the application execution process, the user might change the deadline for completing the submitted task. To deal with the deadline constraint request, application processing would be done in multiple instances that have higher processing capability. The cost of processing of those higher capacity nodes is definitely higher; thus, the user needs to pay more. However, we would manage resource provisioning in a way that the user should not pay unnecessarily.

All notations used throughout this paper are listed in Table \ref{tabnota}.  Assuming that a user submits a task at $t_i$, at the initial stage and then at any event during the processing of the task, the user requests to complete the task within $T_d$ time. Let, $t_j$ be the amount of completed task; hence, the Fog resource would complete the task $t_n$ in $T_d$ time, where $t_n=t_i-t_j$ and it would satisfy the condition $T_{t_n} \leq T_d$, where $T_{t_n}$ is the time needed to complete the task $t_n$. 

\subsection{Network Model}

CloudSim is based on a conceptual abstraction which does not simulate any real networking entities such as routers and switches but considers latency stored in the latency matrix. Topology description consists of all CloudSim entities, such as data centres, hosts and Cloud brokers stored in BRITE  \cite{medina2001brite} format. Using BRITE topology, the description of the shortest path of all pairs is calculated by the Floyd-Warshall algorithm. This latency matrix is generated once while the CloudSim simulation is initialised. In Fog simulation, the location of the nodes will change over the simulation time; thus, it is challenging to model network behaviour. However, we can calculate the shortest path of all pairs using the Floyd-Warshall algorithm but it is not realistic to calculate all of the shortest paths at one time during the initialisation of the simulation. On the other hand, in the Fog scenario, maximum devices will be connected via wireless connections and also wired connections, in some cases. Hence, we need some models for wired and wireless network connections. For wired and wireless connections, ethernet and Wireless Fidelity (WiFi) are not the only network connection types at the present time. Besides ethernet and WiFi, there are serial, fibre, cellular and other types of connectivity. We need some models to implement various types of connectivity.

In Fog computing, network traffic load is unpredictable and latency is crucial in such computing environments in which network latency takes place. Thus, detailed modelling becomes necessary.

In the Fog environment, most of the devices are connected via the wireless or cellular connection, yet some devices could be connected via LAN, WPAN, WAN or Broadband connections. The throughput of the network varies based on the network connectivity type. Generally, network throughput is lower in wireless networks compared with wired networks. If two media are connected with two different linked bandwidths, then the lower bandwidth will be considered as being the available bandwidth  \cite{aceto2012unified}. Therefore, the bandwidth of a link can be defined as:

\begin{equation}
\label{eq1}
B_{WL}= \min_{i=C_A,C_B} b_{w_i} 
\end{equation}

In equation (\ref{eq1}), $B_{WL}$ is denoted as the bandwidth of a link which is the minimum bandwidth of the ports of nodes, $C_A$ and $C_B$. The bandwidth of a path connected via several links and nodes is defined as: 

\begin{equation}
\label{eq2}
B_{WP}= \min_{i=C_A,C_B ..... C_{N}} b_{w_i} 
\end{equation}

Where, $B_{WP}$ represents the bandwidth of a path which is the minimum bandwidth of the ports of nodes $C_A$ and $C_B$....................... $C_{N}$, bandwidth will be further allocated using the Max-Min fairness policy. This policy allows using minimum bandwidth between all links while competing with other users for the bandwidth of the same link. Thus, the available bandwidth of a link or the path for a user will be as follows:

\begin{equation}
\begin{aligned}
\label{eq3}
AB_{WL}= \min_{i=C_A,C_B} b_{w_i}/{N_u}  
or, AB_{WP}= \min_{i=1,2 ..... n} AB_{WL}(i)  
\end{aligned}
\end{equation}

For a link, the bandwidth of the link will be divided by the total number of users, $N_u$, competing for the same bandwidth. In the case of the path, the available bandwidth will be followed by the Max-Min policy where $i$ represents the specifically allocated bandwidth of a link for a specific user.  

The available bandwidth is further varied by the throughput of the link or path. Under high offered loads, the throughputs of the link are measured as 36\%, 71\% and 83\% for the packet size of 64, 512, and 1500 bytes, respectively \cite{wang1999efficient} where the maximum propagation delay is set by 30$\mu s$. According to Wang and Keshav \cite{wang1999efficient}, ``the link throughput is a monotonically increasing and piecewise linear function of the link load and mean packet size''. Xiao and Keshav \cite{xiao2002throughput} defined 1$\mu s$ propagation delay for 802.11a and 802.11b wireless network. Jun et al. \cite{jun2003theoretical} showed that the theoretical maximum throughput is 55\% when CSMA/CA is used, and it is 41\% when the data rate is 11 Mbps and RTS/CTS is used for 1500 byes MSDU size with the absence of transmission errors. From the above description, it is clear that throughput is varied by connection type and packet size. If we consider throughput, then the available bandwidth for link and path will be varied according to their average throughput. Bandwidth is also varied based on the medium type and length \cite{tanenbaum1996computer}. Hence, the availability of a link or a path for a user will be:

\begin{equation}
\begin{split}
\label{eq4}
AB_{WL}= \min_{i=C_A,C_B} b_{w_i}/{N_u} (M_{th}) 
or, AB_{WP}= \min_{i=1,2 ..... n} AB_{WL}(i) \times M_{th}
\end{split}
\end{equation}

Where, $M_{th}$ is the throughput of the medium in percentage.

Network latency depends on four basic parameters: propagation time, transmission time, queuing time and processing delay \cite{forouzan2006data}. Propagation delay varies based on the type of medium (wired or wireless). \\
The total latency for a link of a single communication unit will be as follows:

\begin{equation}
\begin{aligned}
\label{eq5}
N_{dL}= 2(Q_d+T_d+P_d+PR_d)
\end{aligned}
\end{equation}

Where, $N_{dL}$ represents the network delay of a link, $Q_d$ is the queuing delay, $T_d$ is the transmission delay, $P_d$ is the propagation delay, $PR_d$ is the processing delay.

To calculate network delay of a path, it is necessary to consider all intermediate links and nodes. In that, the network delay for a path will be: 

\begin{equation}
\begin{aligned}
\label{eq6}
N_{dP}= Q_{dp}+T_{dp}+P_{dp}+PR_{dp}
\end{aligned}
\end{equation}

Where $N_{dP}$ represents the network delay of the path. $P_{dp}$ is the propagation delay of all intermediate links and can be represented as:

\begin{equation}
\begin{aligned}
\label{eq7}
P_{dp}=\sum_{n=1}^{k} P_{d_n}
\end{aligned}
\end{equation}

Where $k$ is the total number of intermediate links.

Queuing delay, $Q_{dp}$, transmission delay, $T_dp$ and processing delay, $PR_{dp}$ of all intermediate nodes can be represented as follows:

\begin{equation}
\begin{aligned}
\label{eq8}
Q_{dp}=\sum_{n=1}^{k} Q_{d_n} ;
T_{dp}=\sum_{n=1}^{k} T_{d_n} ;
PR_{dp}=\sum_{n=1}^{k} PR_{d_n} 
\end{aligned}
\end{equation}

Propagation time is the delay period which is needed to transfer a data packet from one point to another through the medium. Hence, the time spent by the data packet on the medium is the actual propagation delay. Wired of the wireless medium have different propagation delays. Coaxial cable or fibre optic links have a delay of approximately 5$\mu$ sec/km and microwave links have a delay of 3$\mu$ sec/km \cite{tanenbaum1996computer}. It could be assumed that propagation delay is negligible in an ethernet network \cite{tanenbaum1996computer}. 

Seattle and Amsterdam are located within an 8,000 km distance from each other and the propagation delay of optical fibre for this distance is 40 ms \cite{tanenbaum1996computer}. For the wireless medium, the length of the medium will be the distance of the station from the base station which effects propagation time.

After processing a data packet, packet sends to the media for transmission. The time needed to send packets to the medium is known as transmission time. \ignore{\textbf{Processing delay}:} Processing delay is denoted as the time needed to process the frame header, including the determination of the frame needed to send and bit-level errors. Processing delay depends on the length of the frame ($L$) and transmission rate ($T_r$) \cite{li2017efficient}. The processing delay can be denoted as:

\begin{equation}
\begin{aligned}
\label{eq9}
PR_d=\frac{L}{T_r} 
\end{aligned}
\end{equation} 

The length of the frame is varied by the physical layer technologies (for example, 802.3, 802.11, and 802.16). The transmission rate is varied by the type of network adapter. The processing delay is almost always constant among four different kinds of delays, except for some special occurrences, because hardware-assisted forwarding is being used in modern switches with respect to link speed \cite{hernandez2007one}. Processing delay is constant when a network device is selected and the length of the frame is fixed. Additionally, for certain paths, due to the dependency on the distance and link capacity, the transmission and propagation delays are also constant. Hence, in reality, the queueing delay is only varied randomly because of the variability in the network. Therefore, it is feasible to investigate the queuing delay thoroughly in simulation, by limiting three other delays as the constant \cite{li2017efficient}. Hence, the delay is calculated using Equation \ref{eq5} and can be rewritten as the following:

\begin{equation}
\begin{aligned}
\label{eq10}
N_{dL}= Q_d+PR_{dp}+\sum_{n=1}^{k} A_n
\end{aligned}
\end{equation}

In the above equation, $A_n$ is the constant for processing and transmission delay, and $k$ is the total number of intermediate links. We are dealing with queuing delays and propagation delays. The delays also depends on packet size and link bandwidth. Based on this, the minimum fixed delay follows a linear function of a packet size and link bandwidth is as follows \cite{choi2007analysis, hohn2004bridging}:

\begin{equation}
\begin{aligned}
\label{eq11}
D^{fixed}(W)= W\sum_{i=1}^{h} 1/C_i+\sum_{i=1}^{h} \delta_i
\end{aligned}
\end{equation}

Where, $D^{fixed}$ is the minimum fixed delay, $W$ is packet size, $h$ is the number of hops, $C_i$ is the link capacity and $\delta_i$ is the propagation delay. But, in an operational network, it is not possible for a single user to utilise the whole bandwidth by a single user. We need to calculate the delay by using the currently available bandwidth for a specific user or packet. Propagation delay is already considered in Equation \ref{eq10}. Considering link bandwidth and packet size, Equation \ref{eq11} can be modified as follows:

\begin{equation}
\begin{aligned}
\label{eq12}
N_{dL}= \left\lceil{W\sum_{i=1}^{h} 1/C_i}\right\rceil*\left( Q_d+PR_{dp}+\sum_{n=1}^{k} A_n\right)
\end{aligned}
\end{equation}

If it is assumed that the service time is zero, then no queue will be formed so, we can ignore the queuing delay. Therefore, path delay can be calculated using the following equation: 

\begin{equation}
\begin{aligned}
\label{eq13}
N_{dL}= \left\lceil{W\sum_{i=1}^{h} 1/C_i}\right\rceil*\left( PR_{dp}+\sum_{n=1}^{k} A_n\right) \\
\end{aligned}
\end{equation}

%
%

\subsection{Pricing Model}

The pricing of the Fog resource is important because of its heterogeneous nature. Different types of nodes and networking devices do the computation and data transfer tasks. Aazam and Huh \cite{aazam2015Fog} pproposed a pricing model for IoT in the Fog computing environment. They defined the pricing differently for new and the old customers. According to their probability model, they provided a discount to the users based on their past behaviour of service usage. However, pricing based on user behaviour is unrealistic because the cost of equipment and the operation costs are more important in defining the pricing. On the other hand, Aazam and Huh \cite{aazam2015Fog} considered how concessions will be given to the users but did not consider how the pricing will be defined. 

Many providers such as IBM, Google, AWS, Microsoft and Alibaba are providing services for the IoT environment. They are each following a different pricing model. IBM defines pricing based on three different plans: lite, standard and advanced security plans. The prices are also varied based on deployment  \cite{IBMIoT}. Google defines their prices for IoT devices by data volume exchanged through the Cloud IoT core \cite{GoogleIoT}. Microsoft defines pricing based on a monthly plan  \cite{MicrosoftIoT}. Alibaba Cloud IoT solution pricing is based on the type of service and a user pay-as-you-go model \cite{AliBabaIoT}.
AWS estimates pricing based on four key items: 1) connectivity charge per minute, 2) messaging charge per message, 3) device shadow and registry charge, and 4) rules engine charge \cite{AmazonIoT}. Table \ref{AWSIoT} shows the pricing of Amazon IoT solutions.

\begin{table}[htbp]
	\centering
	\small
	\caption{AWS IoT pricing}
	\label{AWSIoT}
	\begin{tabular}{|L{5cm}|L{3cm}|} \hline
		\textbf{Services} & \textbf{Price} \\ \hline
		Connectivity (per million minutes) & \$0.08 - \$0.132  \\ \hline
		Messaging (per million messages/5KB) & \$1.00 - \$1.65 \\ \hline
		Device shadow and registry (per million operations/1 KB) & \$1.25-\$1.88 \\ \hline
		Rules engine (per million rules triggered or action executed/5KB)& \$0.15 - \$0.25 \\ \hline
	\end{tabular}
\end{table}

Of them all, the AWS pricing model is the most suitable for the pricing estimation in our simulation since we are evaluating our model in a network perspective. Thus, pricing by connectivity and by messaging is best suited to our simulation environment. But one issue arises: we need to separate the processing price by Fog layer. In our simulation, we assume that the processing will be done in a Fog device, a Fog server and on the Cloud when necessary. Thus, it is obvious that the data processing and price exchange for each level are different. According to the pricing of IBM Waston Internet of Things, they are charging half for edge processing. Using the IBM pricing model, we can assume that the half price is where computation takes place on the Fog server and the price is one-third in the case of the Fog device. Again, the issue is that time-sensitive application users are paying less compared with the users using the Cloud. On the other hand, the infrastructure and maintenance costs of the Cloud are high. To solve this problem, we should prioritise time-sensitive applications in the Fog environment. According to our assumption, we can formulate pricing for connectivity, messaging, registry and rules or action.

\textbf{Connectivity cost:} Cost of connectivity can be calculated using Equation \ref{eqcc}:

\begin{equation}
\label{eqcc}
\begin{split}
C_{total} =  CP ( \sum{ C_c}+\frac{1}{FS_x}\sum{C_{FS}}  +\frac{1}{FD_x}\sum{C_{FD}} ) \times10^{-6}   
\end{split}
\end{equation}

Where, $C_{total}$ is the total charges for connectivity, $CP$ is the unit price for connectivity, $C_c$ is the connection time to the Cloud in minutes, $C_{FD}$ is the connection time to the Fog device in minutes, $C_{FS}$ is the connection time to the Fog server  in minutes, $FS_x$ is the cost variable for the Fog server (two for our case), and $FD_x$ is the cost variable for the Fog device (three for our case).

\textbf{Messaging cost:} Cost of messaging can be calculated using Equation \ref{eqmc}:

\begin{equation}
\label{eqmc}
\begin{split}
M_{total} =  MP(  \sum_{i=1}^{n}{\left\lceil{\frac{{M_c}_i}{U}}\right\rceil}+\frac{1}{FS_x}\sum_{i=1}^{n}{\left\lceil{\frac{{M_{FS}}_i}{U}}\right\rceil} 
+\frac{1}{FD_x}\sum_{i=1}^{n}{\left\lceil{\frac{{M_{FD}}_i}{U}}\right\rceil} ) \times10^{-6}
\end{split}
\end{equation}

In the above equation, $M_{total}$ is the total charges for messaging, $MP$ is the unit price of messaging, $M_c$ is the messaging requests to the Cloud, $M_{FD}$ is the messaging requests to the Fog device, $M_{FS}$ is the messaging requests to the Fog server, $U$ is the unit of data size (5KB in our case), $n $ is the total number of requests, $FS_x$ is the cost variable for the Fog server (two for our case), and $FD_x$ is the cost variable for the Fog device (three for our case).

\textbf{Registry or Shadow Cost:} Cost of registry or shadow can be calculated using Equation \ref{eqrsc}:

\begin{equation}
\label{eqrsc}
\begin{split}
S_{total} =  SP (  \sum{SR_c}+\sum{\frac{1}{FS_x}SR_{FS}}  +\sum{\frac{1}{FD_x}SR_{FD}} ) \times10^{-6}  
\end{split}
\end{equation}

Where, $S_{total}$ is the total charge for shadow and registry. $SP$ is the unit price of registry or shadow cost, $SR_c$ is the registry or shadow requests to the Cloud in KB, $SR_{FD}$ is the registry or shadow requests to the Fog device in KB, $SR_{FS}$ is the registry or shadow requests to the Fog server in KB, $FS_x$ is the cost variable for the Fog server (two for our case), 
$FD_x$ is the cost variable for the Fog device (three for our case).

\textbf{Processing Cost:} Cost of processing can be calculated using Equation \ref{eqpc}:

\begin{equation}
\label{eqpc}
\begin{split}
P_{total} =  PP(  \sum_{i=1}^{n}{\left\lceil{\frac{{P_c}_i}{U}}\right\rceil}+\frac{1}{FS_x}\sum_{i=1}^{n}{\left\lceil{\frac{{P_{FS}}_i}{U}}\right\rceil} 
+\frac{1}{FD_x}\sum_{i=1}^{n}{\left\lceil{\frac{{P_{FD}}_i}{U}}\right\rceil} ) \times10^{-6}
\end{split}
\end{equation}

In the above equation, $P_{total}$ is the total charges for the rules engine, $PP$ is the unit price for the processing, $P_c$ is the processing request to the Cloud, $P_{FD}$ is the processing requests to the Fog device, $P_{FS}$ is the processing requests to the Fog server, $U$ is the unit of data size (5KB in our case), $n$ is the total number of requests, $FS_x$ is the cost variable for the Fog server (two for our case), and $FD_x$ is the cost variable for the Fog device (three for our case).

In our simulation, we only considered the connectivity, processing and messaging. Shadow and registry did not take place; hence, the total cost per application will be as follows:

\begin{equation}
\begin{split}
AT_{cost}=C_{total}  + M_{total}  + P_{total} \\
\end{split}
\end{equation}

\subsection{Proposed Policy}
\label{pro_pol}
As the resources are dynamic in the Fog environment, we employ priority-based techniques to select appropriate resources by considering detail characteristics of the Fog resources. We need to undertake the following Fog related assumptions to select appropriate resources:

\begin{itemize}
\item Distance (geographical location of Fog devices)
\item Utilisation (free resources)
\item Response time (including migration time and execution time)
\item Availability (due to the of battery life of the Fog devices)
\item Fluctuating behaviour of CPU availability 
\end{itemize}

\begin{figure}[htbp]
	\centering
	\includegraphics[width=4.5in]{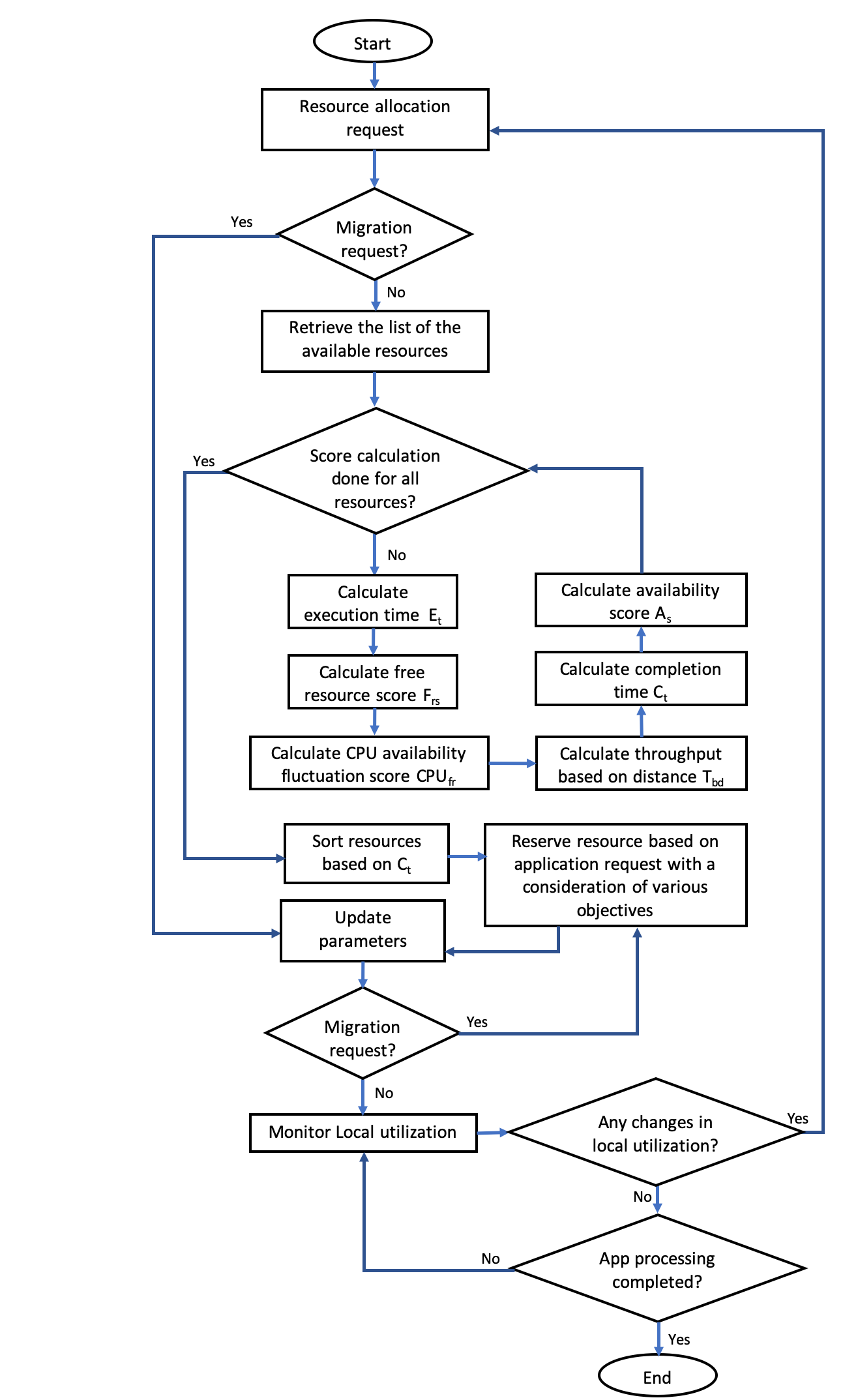}
	\caption{Resource allocation in Fog infrastructure.}
	\label{algoflow1}
\end{figure} 

Fog devices are not exclusively dedicated for Fog application processing. Most of these devices are mainly responsible for native application processing. If the participating Fog devices have some currently available computation power, they can use them for Fog processing. Even during Fog application processing, CPU utilisation might vary, based on the load of the native applications. This can be referred to as fluctuating behaviour of CPU availability. The rate of fluctuation in CPU availability during a Fog application execution can be referred as CPU availability fluctuation rate.
In a Fog environment, such a situation might arise when the users' requirements change dynamically over time. The application broker should be able to schedule tasks in a way that the application could serve the requests of the user. Hence, some resources should be reserved for tentative future application requests. We assume that Fog application services will not responsible for the bulk processing, in other words, long-term processing. Thus, the reservation will be updated from time to time based on user requests to the application. During the resource allocation and scheduling process, the proposed Multi-Criteria-based (MC-based) policy will be able to cope with basic Fog computing characteristics like limited resource capacity, dynamic resource availability and mobility. We assume that there will be no case when the Fog resource is not sufficient to complete the service. In other word, we do not have available resources to serve for the requested application. For example, we need 1000 MIPS pressing power for a certain time to complete an application but we do not have such resources available in the Fog environment. We also assume that there will be no case when Fog resources will be completely full and the user's request needs to be completed by the Cloud resources. That is, all of the time-sensitive processing will be done in the Fog environment by utilising Fog infrastructure. Of course, the system is depending on the Cloud for long-term storage and bulk processing which are not time-sensitive. A sensor application will be run on each Fog device to monitor the resources which will be handled on the Fog platform. The Fog server will have all the resources and information available on the Fog devices that are connected within that Fog server. Resource allocation and scheduling will be completed with the following steps:        
\begin{enumerate}
\item Retrieve the list of available resources.
\item Find appropriate resources considering response time, utilisation, availability, location, and CPU availability.
\item Allocate resources based on objective based resource allocation. 
\item Reserve resources based on the application request with a consideration of various objectives.
\item Monitor the local resource usage and reassign the task when necessary.
\end{enumerate}
The steps of the proposed algorithm are shown in Fig. \ref{algoflow1}. The proposed MC-based resource allocation algorithm is presented in Algorithm \ref{raalg}.

\begin{algorithm}[!t]
	\caption{MC-based Resource allocation for Fog application.}
	\label{raalg}
	\small
	\hspace*{\algorithmicindent} \textbf{Input: $ResourceList<CPU_{cap}, Free_{res},Avail,Thro_{bod},CPU_{avalfl}>, App_{req}, Usr_{req}$}  \\
	\hspace*{\algorithmicindent} \textbf{Output:$ResouceList<rid>$} 
	\begin{algorithmic} 
		\IF {$ResourceList[].size \neq null$ \AND $request \neq migration_{req}$}
		\FORALL{$ResourceList[]$}
		\STATE $E_t \leftarrow Calculate\: execution \:time$
        \STATE $Frs_{s} \leftarrow Calculate \:free resource \:score$
        \STATE $CAF_s \leftarrow Calculate \:CPU \:availability \:fluctuation \:score$
        \STATE $T_{bd} \leftarrow Calculate \:throughput\: based\: on \:distance$
        \STATE $C_t \leftarrow Calculate \:Completion \:time$
        \STATE $A_s \leftarrow Calculate\: availability\: score$
		\ENDFOR
        \STATE $Sort\: by\: C_t$
        \RETURN $ResourceList <rid>$ 
        \ELSIF {$request = migration_{req}$}
        \STATE Update parameters
		\STATE Reserve resource based on previous application request
        \STATE Find appropriate resource for migration
        \STATE $Sort\: by\: C_t$
        \RETURN $ResourceList <rid>$  
%
		
		\ELSE \RETURN $NULL$
		\ENDIF
	\end{algorithmic}
\end{algorithm}
Assume that there are five devices which have enough available resources to complete a task. Each device has different migration time, execution time, free CPU, availability, distance and CPU availability fluctuation as presented in Table \ref{devpara}. Assume that we need to allocate 1000 Units (1000 MIPS) jobs and the user has deadline requirements which are five units of time in which to complete the job.

\begin{table*}[ht]
	\centering
    \scriptsize
	\caption{Different parameters for multiple Fog devices.}
	\label{devpara}
	\begin{tabular}{L{1cm}L{1cm}C{0.30cm}C{0.25cm}C{0.30cm}C{0.25cm}C{0.30cm}C{0.25cm}C{0.30cm}C{0.25cm}C{1.2cm}C{0.7cm}C{0.9cm}C{0.6cm}C{0.6cm}} 
		\toprule
\multirow{2}{*}{\textbf{Devices}} & \textbf{CPU Capacity (MIPS)} & \multicolumn{9}{c}{\textbf{Response Time}} &  \multirow{2}{*}{\textbf{$F_{rs}$}} & \multirow{2}{*}{\textbf{Availability}} & \multirow{2}{*}{\textbf{$T_{bd}$}} & \multirow{2}{*}{\textbf{$C_{af}$}}  \\ \cline{3-11}
&&\multicolumn{8}{c}{\textbf{Migration Time}}&\textbf{Execution Time}&&&& \\ \hline 
FD1	& 1000 & D2 &	2 & D3 & 4 & D4 & 3 & D5 & 1 & 1 & 50\% & 10 Sec & 0.9 & 50\%  \\ \hline
FD2	& 500 &	D1	& 2	& D3 &	5 &	D4 &	2 &	D5 &	3 &	2 &	60\% &	12 Sec &	0.8	& 80\% \\ \hline
FD3	& 100 &	D1	& 4	& D2 &	5 &	D4 &	4 &	D5 &	2 &	10 &	30\% &	20 Sec &	0.5	& 100\% \\ \hline
FD4	& 200 &	D1	& 3	& D2 &	2 &	D3 &	4 &	D5 &	1 &	5 &	40\%	& 30 Sec &	0.7	& 130\% \\ \hline
FD5	& 300 &	D1	& 1	& D2 &	3 &	D3 &	2 &	D4 &	1 &	3.33	& 20\% &	5 Sec	& 0.55	& 90\%  \\ \hline
\bottomrule
\end{tabular}
\end{table*}


Now the question is: which resource would be more suitable for the job as stated above? All devices have the capability of completing the requested jobs, except the third Fog device. To deal with this, our algorithm will choose the best suitable resource to complete the job. Initially, the algorithm will ignore the migration time and find the time that will be needed by each device in order to complete the task. Table \ref{ctas} shows the outcome of the completion time calculation.

\begin{table*}[ht]
	\centering
    \footnotesize
	\caption{Completion time and availability score}
	\label{ctas}
	\begin{tabular}{L{1.2cm}C{1.5cm}C{1.5cm}C{1.5cm}C{1.5cm}C{1.5cm}C{1.5cm}} 
		\toprule
Devices	& $E_t$	& $Frs_s$ & $CAF_s$ & $T_{bd}$ & $C_t$ & $A_s$ \\ \hline
FD1	&	1	&	0.5	&	0.5	&	0.9	&	4.44	&	2.25	\\ \hline
FD2	&	2	&	0.6	&	0.8	&	0.8	&	5.21	&	2.304	\\ \hline
FD3	&	10	&	0.3	&	1	&	0.5	&	66.67	&	0.3	\\ \hline
FD4	&	0.5	&	0.4	&	1.3	&	0.7	&	1.37	&	21.84	\\ \hline
FD5	&	0.33	&	0.2	&	0.9	&	0.55	&	3.37	&	1.485 \\ 
\bottomrule
\end{tabular}
\end{table*}


Availability score and completion time will be calculated using the following equations:

\begin{equation}
\label{eq100}
A_s= \frac{E_t}{F_{r} \times CAF_{s} \times T_{bd}} 
\end{equation}

Where, $F_{rs}$ is Free resources score, $CAF_{s}$ is CPU availability fluctuation score and $T_{bd}$ is Throughput based on distance.

\begin{equation}
\label{eq101}
C_t= \frac{A_v}{A_s} 
\end{equation}

Where $A_v$ is the Availability and $C_t$ is the completion time.

According to Table \ref{devpara}, if we want to execute the 1000 MIPS task on FD1 then it will take two seconds to complete the task because 50\% of the CPU resource is available. Again, based on the CPU fluctuation rate, we can assume that the completion of the task might take four seconds. Finally, based on throughput, the system will take 4.44 seconds to complete the task by FD1.

We are assuming that Fog devices are dynamic so the allocation policy will assign the job to a system which will take less time to complete the job. We can assign the job to the best fit system but in such a case, if Fog device utilisation for native application becomes higher in a way that the Fog application cannot be run, then there is a need to migrate the jobs to another system. We need to schedule tasks in such a way that they can migrate and be reassigned to another system to minimise SLA violation. Due to this, FD4 is the most suitable resource to complete the submitted task as shown in Table \ref{ctas}.

If anything changes on the job execution (which occurs from the changing behaviour of users or native resource utilisation), this needs to be handled during the job execution. Then the jobs need to be migrated to another device. Hence, it is necessary to consider migration time. In such cases, the job will be migrated to the system which will take less time to migrate the job and with high availability. In the case of the above-stated job, if FD4 native utilisation restricts the execution of the Fog application, then the system will not choose FD5 for job migration; it will choose FD1 because FD1 has high availability score with the closest completion time compared with FD5.   

More details about response time, free resources, availability, throughput and CPU availability are discussed in the following:

Response time, $R_t$ depends on migration time, $M_t$ and execution time, $E_t$ as shown in Equation \ref{eq102}. Response time indicates the time required to get a response after completing a particular task. This includies sending the task, execution and returning the results to the sender.

\begin{equation}
\label{eq102}
R_t= M_t + E_t + N_{dL}
\end{equation}

Migration time depends on data size, $D_s$, bandwidth, $b_w$ and throughput, $t_h$ as shown in Equation \ref{eq103}:

\begin{equation}
\label{eq103}
M_t= \frac{D_s}{B_w\times T_h} 
\end{equation}

Execution time depends on the size of the jobs in MIPS, $J_s$ and CPU speed in MIPS, $CPU_s$ as shown in Equation \ref{eq104}:

\begin{equation}
\label{eq104}
E_t= \frac{J_s}{CPU_s} 
\end{equation}

Free resources depend on the current utilisation of the CPU. It will always vary on the applications running on the device itself (native applications).

Availability depends on the battery life of the mobile device. If the device is directly connected to power, with an appropriate backup, then there is no downtime and the availability is 100\%. Battery power is non-linear based on the profile of battery discharge \cite{yurur2015modeling}. We assume that the Fog device will keep updating availability and the Fog platform will handle this. However, it can be calculated as follows: Assume that a mobile device has 60\% charge left and three applications are running, with App1, App2 and App3 discharging battery power of 0.5\%, 0.2\%, and 0.3\%, respectively in every minute. Hence, the availability can be calculated using Equation \ref{eq105}:

\begin{equation}
\label{eq105}
A_v= \frac{A_b}{\sum_{i=1}^{n} A_{dr_i}} 
\end{equation}


Throughput always depends on the distance of wireless connections. Throughput is always decreasing while distance is increasing \footnote{\url{https://www.geckoandfly.com/10041/wireless-wifi-802-11-abgn-router-range-and-distance-comparison/}}. Hence, throughput depends on geographical distance, $G_d$ and maximum supported distance by the device, $SD_{max}$ as shown in Equation \ref{eq106}:

\begin{equation}
\label{eq106}
t_h= \frac{G_d}{SD_{max}} 
\end{equation}


CPU availability is the free CPU that can be utilised over time. Due to the nature of the Fog device, the fluctuation of CPU availability will be high. We will need to calculate the rate of fluctuation. Let us assume that in different time frames, T1, T2, T3, T4, and T5, the available CPU was 10\%, 20\%, 5\%, 30\% and 20\%, respectively. Compared with T1 $\rightarrow$ T2, fluctuation is 100\%. Similarly, for T2 $\rightarrow$ T3, it is 75\%; for T3 $\rightarrow$ T4, it is 500\%; for T4 $\rightarrow$ T5, it is 33\%. Thus, the average fluctuation rate can be calculated using Equation \ref{eq107}, where, 
$CPU_{fr}$ is the average CPU fluctuation rate and $Fr_r$ is the fluctuation rate in each time-frame. 

\begin{equation}
\label{eq107}
CPU_{fr}= \frac{\sum_{i=1}^{n} A_{Fr_i}}{n} 
\end{equation}

\begin{figure}[htbp]
	\centering
	\includegraphics[width=2.5in]{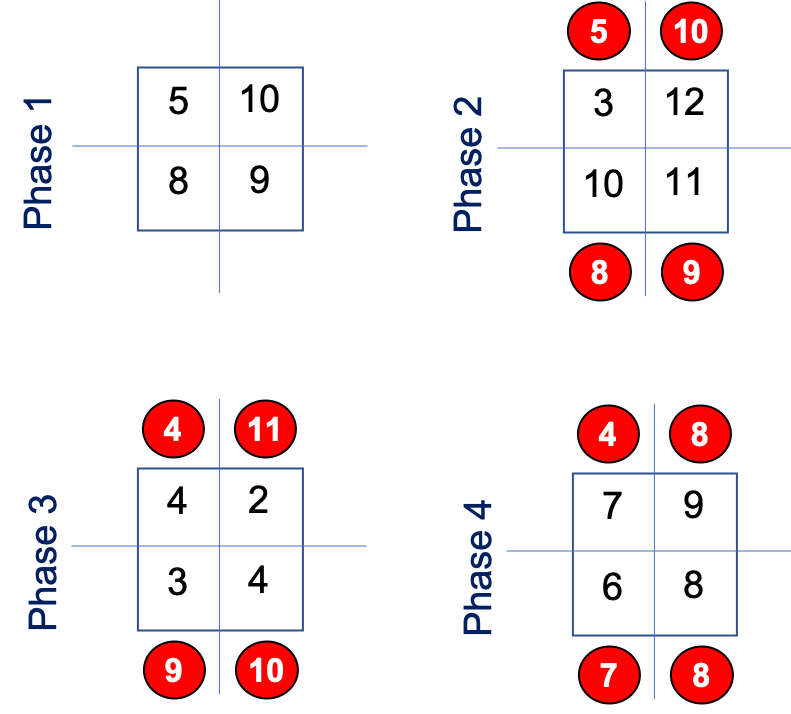}
	\caption{Resource reservation in Fog infrastructure.}
	\label{reservation}
\end{figure}

\textbf{The algorithm for resource reservation:}
The resource will be reserved based on a the previous history of task submission. For this purpose, the system will store the number of application requests in a specific time period, with the average size of the application. Then, the system will reserve the resources according to the number of application requests and the average size of the applications. For instance, if five application requests are coming in a specific period of time and the size of each application is 20 units, then the system will reserve the resources that can execute 100 units of applications. Now, the question is: which device will reserve the resources? In a Fog cluster, the system will reserve the resources based on the volume of users' requests going to that cluster. Generally, Fog devices will always accept the requests from peer Fog devices. But, after the addition of resource reservation, the system will only accept the request if the Fog cluster has available space after reservation, in order to handle dynamic user behaviour. Fig. \ref{reservation} shows how reservation will work in the system scenario. According to this figure, the reservation is shown in the red circle. Initially, none of the resources will be reserved for Fog applications but, from the second phase, resources will be reserved based on the previous application requests. In each rotation of time, the reservation value will be updated.   

Since it is based on previous task submission history, the system will perform better over time. By following this idea, we can simulate different slices of time rotation. Users of the system can choose different time slices according to their needs. Algorithm \ref{resres} shows the steps for calculating and updating the value of resource reservation.

\begin{algorithm}[htbp]
	\caption{Resource reservation for Fog application.}
	\label{resres}
	\hspace*{\algorithmicindent} \textbf{Input: $ ResourceList<FogDevice>, CurrentUtil<FogDeviceId,Utilisation>$}  \\
	\hspace*{\algorithmicindent} \textbf{Output:$Status$} 
	\begin{algorithmic} 
		\IF {$ResourceList[].size \neq null$ }
		\FORALL{$ResourceList[]$}
		\STATE $id \leftarrow FogDevice[i].FogDeviceId$
		\STATE $CU_z \leftarrow CurrentUtil[id].Utilisation$
		\STATE $R_v \leftarrow FogDevice[i].ResValue$
		\STATE $L_{AR} \leftarrow FogDevice[i].LastAppReq$
		\STATE $T_{AP} \leftarrow FogDevice[i].TotalAppProcessed$
		\STATE $Req_{res} \leftarrow (R_v+L_{AR})/T_{AP}$
		\STATE $FogDevice[i].Utilisation \leftarrow CU_z + Req_{res}$
    	\ENDFOR
    	\ELSE
    	\RETURN $Reservation Failed$
     	\ENDIF
     	\RETURN $Success$
     	
	\end{algorithmic}
\end{algorithm}

Algorithm \ref{resres} takes the Fog device list and the current utilisation of each Fog device as input. Then the required reservation ($Req_{res}$) is calculated using the reservation value ($R_V$), utilisation of the last application request ($L_{AR}$) and the total application processed ($T_{AP}$) by the Fog device. Finally, Fog device utilisation is updated to the current utilisation ($CU_z$) and required reservation ($Req_{res}$).  Based on the success or failure of resources reservation, the algorithm generates the outcome.

\textbf{The algorithm for handling dynamic users' behaviour:}
When the dynamic changes occur in the system environment, the system will call Algorithm \ref{dynbeh} which will find an appropriate resource for the submitted application. Here we consider a deadline as dynamic user behaviour. Algorithm \ref{dynbeh} will be triggered if users want to have their application outcome in less time, compared with the submitted application request. Here, task execution time and task migration time are both considered before selecting resources chosen by the MC-based resource allocation algorithm. If this algorithm returns as null, the system will look for an appropriate resource, using the MC-based resource allocation algorithm. Once the system finds an appropriate resource, which will meet the deadline by matching both execution time and migration time, then the application will be submitted to the resource that has been chosen.

\begin{algorithm}[htbp]
	\caption{Handling dynamic user behaviour in Fog environment.}
	\label{dynbeh}
	\hspace*{\algorithmicindent} \textbf{Input:  $ResourceList<FogDevice>$,$Usr_{req}$}  \\
	\begin{algorithmic} 
		\STATE $TR_{list}<FogDevice>$= MC-basedAlgo<$ResourceList<FogDevice>$,$Usr_{req}$> 
			\FORALL{$TR_{list}[]$}
			\STATE $TE_T \leftarrow TR_{list}[i].TaskExeTime$
			\IF {$TE_T$ < $Usr_{req}$+$MigrationTime$}
			\STATE $id \leftarrow TR_{list}[i].id$
			\STATE $Break$
			\ENDIF
			\ENDFOR
		    \STATE MC-basedAlgo<$id$,$App_{req=migration}$> 
	\end{algorithmic}
\end{algorithm}

Algorithm \ref{dynbeh} is taking the Fog device list and user requests deadline as input. Then it is finding all the tentative resource lists using Algorithm \ref{raalg} in which application execution can be performed within the changed deadline requested by the user. Then it is finding a resource in which task execution time is less than the summation of user-requested time and migration time. Finally, it is calling Algorithm \ref{raalg} to perform the migration.

\subsection{Performance Metrics}
Several metrics such as delay, completion time (processing time), SLA violation, and cost have been chosen to evaluate the performance of the proposed MC-Aware policy. The aim of Fog computing is to reduce delay and processing time, in order to serve time-sensitive applications. Hence, these two metrics are appropriate for performance evaluation. However, cost and SLA violation should not be very high, while delay and completion time should be less. Therefore, we need to measure costs and SLA violations to evaluate the effectiveness of the system.

\subsubsection{Completion time}
Most of the computation tasks are intended to be done by the Fog device in the Fog computing environment. However, based on user requirements, we need to send some requests to the Cloud for processing or storage. Therefore, the average completion time of each request depends on the communication between users with the Fog device, Fog device with Fog server, Fog server with the Cloud, and vice versa. The average completion time of each request also depends on the processing delay of the device itself. If any Fog device or Fog server is unable to process the request due to insufficient resource availability, then they will forward the request to the closely located peer Fog device or Fog server. We can ignore those communication delays since they are near to the sender and delay is minimal. We will consider the delay between the user to FD, FD to FS, FS to the Cloud and processing time needed by each component. Let $P_u$ be is the total number of communications sent by the user, more precisely the total number of packets. $PC_{u}$ is the number of packets required to be sent to the Cloud. The total number of packet transmissions can be calculated by as follows:

\begin{equation}
\begin{split}
TP = P_u + PC_u + PC^r_u 
+ (P_u-PC_u)^r + PC^r_u
\end{split}
\end{equation}

Let $tP_u$ be the total time required to send $P_u$ packet. Similarly, $tPC_u$ is denoted as the total time required to send $PC_u$ packets. From the above equation, the total delay of the $P_u$ packet and the average delay of each packet can be calculated as follows:

\begin{equation}
\begin{split}
DP_{total} =tP_u + tPC_u + tPC^r_u 
+ (tP_u-tPC_u)^r + tPC^r_u
\end{split}
\end{equation}

\begin{equation}
\begin{split}
DP_{avg} = \frac{tP_u + tPC_u + tPC^r_u + (tP_u-tPC_u)^r + tPC^r_u}{P_u}
\end{split}
\end{equation}

$P_{ip}$ is the total number of communications needed for internal processing by the Fog for a user request. It can denoted as follows:

\begin{equation}
TIP = P_{ip} + P^r_{ip} 
\end{equation}

From the above equation, the total and the average delay of each internal communication can be calculated by as follows:

\begin{equation}
DIP_{total} = tP_{ip}^{Fog} + tP^{rFog}_{ip} + tP_{ip}^{Cloud} + tP^{rCloud}_{ip} 
\end{equation}

\begin{equation}
DIP_{avg} = \frac{tP_{ip}^{Fog} + tP^{rFog}_{ip} + tP_{ip}^{Cloud} + tP^{rCloud}_{ip} }{P_{ip}^{Fog}+P_{ip}^{Cloud}} 
\end{equation}

$tP_{fd}$, $tP_{fs}$ and $tP_{c}$ are the processing times required by the Fog device, Fog server and Cloud respectively. So, the total processing time will be as follows:

\begin{equation}
TPT =  tP_{fd} + tP_{fs} + tP_{c}
\end{equation}

The average completion time of a single user request will be as follows:


	\begin{equation}
	\begin{split}
	CTU_{avg} = \{(tP_u + tPC_u + tPC^r_u + (tP_u-tPC_u)^r  + tPC^r_u)+ (tP_{ip}^{Fog} + tP^{rFog}_{ip} + tP_{ip}^{Cloud} \\ + tP^{rCloud}_{ip}) +(tP_{fd} + tP_{fs} + tP_{c})\}/P_u 
	\end{split}
	\end{equation}

Total completion time for an application instance with the number of the request, $n$, would be as follows:

\begin{equation}
\begin{split}
CTA =  \sum_{k=1}^{n} (DP_{total})_k + \sum_{k=1}^{n} (DIP_{total})_k  + \sum_{k=1}^{n} TPT_k
\end{split}
\end{equation}

Average completion time of the system for each request will be:

\begin{equation}
CTA_{avg} =  \frac{\sum_{k=1}^{m} CTA}{\sum_{k=1}^{m} n_k}
\end{equation}

\subsubsection{Cost}
In the proposed system, the user application is used in the Fog system resources as well as Cloud system resources. Therefore, the user has to pay for both systems. We can only consider the processing cost in the Fog computing environment. From Equation 29, the cost for using Cloud and Fog environments for a single request will be as follows:

\begin{equation}
	\begin{split}
	TC_{req} =  \{(DP_{total}+DIP_{total}+TPT) \times C_{Fog}\}  + \{(tPC_u + tPC^r_u+tP_{ip}^{Cloud} \\ + tP^{rCloud}_{ip} +tP_c) \times C_{Cloud}\}
	\end{split}
\end{equation}

\begin{equation}
	\begin{split}
	TC_{req} =  \{(DP_{total}+DIP_{total}+TPT) \times AT_{cost}\}  + \{(tPC_u + tPC^r_u+tP_{ip}^{Cloud} \\ + tP^{rCloud}_{ip} +tP_c) \times AT_{cost}\}
	\end{split}
\end{equation}

In the above equation, $C_{Fog}$ is denoted as the unit cost of the Fog environment and $C_{Cloud}$ denoted as the unit cost of Cloud resources. The total cost for an application instance with $n$, number of the request would be as follows:

\begin{equation}
TC =  \sum_{k=1}^{n} (TC_{req})_k 
\end{equation}

\subsubsection{Service Level Agreement (SLA)}

Service quality is generally guaranteed by the SLA. The provider is responsible for the maintenance of an adequate response time to avoid the violation of SLA. We will measure the response time and cost as agreed by the SLA. In the Fog, users' dynamic requirements will be response time or cost. If the provider is unable to serve according to the agreed requests of the users, then the provider will have to pay for the violation. An SLA violation penalty will follow a linear function which is similar to other related works of \cite{yeo2005service,rana2008managing,irwin2004balancing,wu2011sla,wu2014sla}. The function will be as follows:

 
\begin{equation}
Penalty = \alpha+\beta \times DT 
\end{equation}
Where, $\alpha$ is a constant value for the penalty, $\beta$ is the penalty rate and $DT$ is the delay time. Delay time is the extra time that users waited as stated in SLA for obtaining a response. The percentage of SLA violation is also calculated.

\section{Experimental Setup and Simulation Parameters}
\subsection{Experimental Setup}
The evaluation of managing user dynamic behaviour for time-sensitive application handling in the Fog is carried out in a simulation environment. It is very difficult to develop a real, controlled environment for the experiments. Therefore, we choose simulation for our experiments. CloudSim simulator \cite{calheiros2011cloudsim} is used to simulate the Fog environment. We added Fog devices and a Fog server by including their individual features, such as network connectivity, how far they are located from the access points and how much battery power they have, keeping all other existing features that one host has in CloudSim. The varying number of application submissions is the experimental procedure by which the performance of the proposed method is observed. A synthetic workload is used since the real workload of the Fog environment is not currently available  \cite{mahmud2018cloud,mahmud2018quality}. Based on the previous literature \cite{mahmud2018quality,skarlat2017towards}, we tested the proposed method by increasing the number of application submissions. Hence, 70 to 560 applications have been submitted to the Fog environment, increasing by 70 applications each time.

\subsection{Simulation Parameters}
Table \ref{tsFogSimPara} illustrates the parameters used for the simulation. Table \ref{DybFogSimPara2} represents the other parameters that are used to model dynamic user behaviour, distance, battery life, and CPU availability fluctuations.

\begin{table}[htbp]
	\centering
	\footnotesize
	\caption{Simulation Parameters}
	\label{tsFogSimPara}
	\begin{tabular}{L{5cm}|L{2.5cm}} 
		\toprule[1pt]
		\textbf{Parameter} & \textbf{Value} \\ \hline
		
        \textbf{Fog Server Configuration} & \\ \hline
		MIPS (Millions Instruction Per Second) & 10000  \\ \hline
		No of Pes & 1 \\ \hline
		No of Host & 1 \\ \hline
		Bandwidth (bps) & 1000000 \\ \hline
		RAM & 302768 \\ \hline
		
		\textbf{Fog Device Configuration} & \\ \hline
		MIPS (Millions Instruction Per Second) & 2000 to 6000  \\ \hline
		No of Pes & 1 \\ \hline
		No of Host & 1 \\ \hline
		Bandwidth (bps) & 100000 \\ \hline
		RAM & 2048 \\ \hline
		
		\textbf{Task Configuration} & \\ \hline
		Task Length (MI) & 3000  \\ \hline
		Data Size & 5120 and above \\ \hline
		
		\textbf{Sub Task Configuration} & \\ \hline
		Task Length (MI) & 500  \\ \hline
		Data Size & 5120 and above \\ \hline
		\bottomrule[1pt]
		
	\end{tabular}
\end{table}

\begin{table}[htbp]
	\centering
	\footnotesize
	\caption{Other Parameters}
	\label{DybFogSimPara2}
	\begin{tabular}{L{5cm}|L{2.5cm}} 
		\toprule[1pt]
		\textbf{Parameter} & \textbf{Value} \\ \hline
		\textbf{No of Task per App} & 10\\ \hline
		\textbf{Minimum deadline for tasks} & 4\\ \hline
		\textbf{CPU availability fluctuation} & 50\% - 130\%\\ \hline
		\textbf{Distance} & 5 to 40 Meter\\ \hline
		\textbf{Battery power} & 20\% to 90\% \\\hline
        \textbf{CPU Utilisation variation during task execution} & 10\% to 40\%         \\
        \bottomrule[1pt]
		
	\end{tabular}
\end{table}


The main goal of this work is to allocate the application tasks to the Fog infrastructure, and not to the Cloud. During the simulation, different evaluation scenarios were followed. In the first evaluation scenario, 70 applications were submitted to the Fog environment in the initial stage, then the number of application submission was increased gradually, up to 560 applications. We measured delay, processing time and cost for this evaluation scenario. In the second evaluation scenario, the increasing number of application submission SLA-violation was measured with and without reservation. The last evaluation scenario was a of each dynamic parameter for users and devices. These dynamic parameters were variations of user deadlines, free resource variation, battery power variation and CPU utilisation fluctuation variation. 

\section{Results and Discussion}
The proposed MC-based policy is compared with recently proposed QoE-aware \cite{mahmud2018quality} policy. This is because QoE aware application placement policy performed better, compared with three other recently proposed policies: MeFoRE \cite{aazam2016mefore}, QoS-aware \cite{skarlat2017towards}, and CloudFog \cite{lin2015cloud}. We tested our proposed and QoE-aware algorithm in a simulated environment by employing with and without resource reservation. In addition, we measured the Service Level Agreement (SLA) violation since we are considering that the submitted application by the user and device are both dynamic.

We evaluate the proposed MC-based policy in two different experimental settings. In the first setting, we fixed user and resource dynamic behaviour parameters within the range presented in Table \ref{DybFogSimPara2}. In a different setting, the variation of dynamic behaviour parameters was considered.  


\subsection{Dynamic Behaviour Parameters Variation Within a Range}
\begin{figure}
    \centering
    \begin{subfigure}[b]{0.495\textwidth}
        \frame{\includegraphics[width=\textwidth]{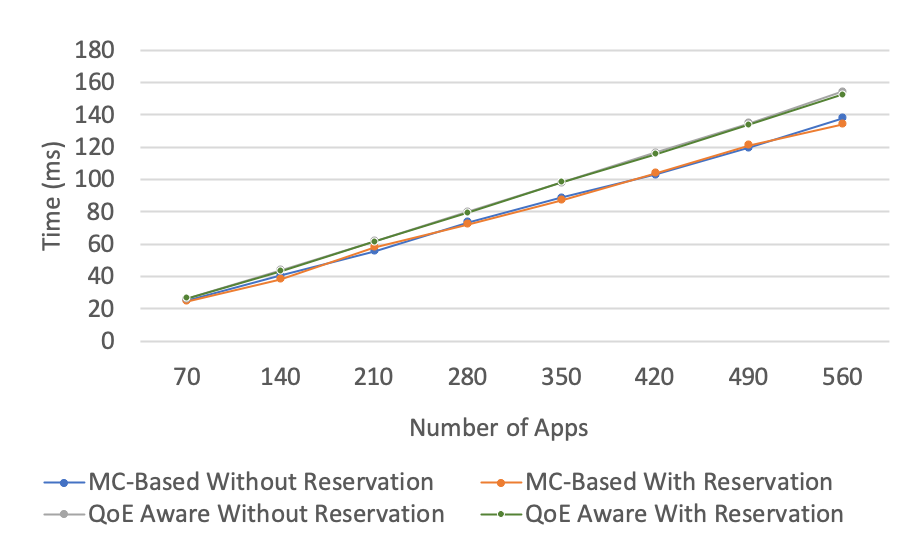}}
        \caption{Average delay}
        \label{1AvgDelay}
    \end{subfigure}
    ~ 
    \begin{subfigure}[b]{0.495\textwidth}
        \frame{\includegraphics[width=\textwidth]{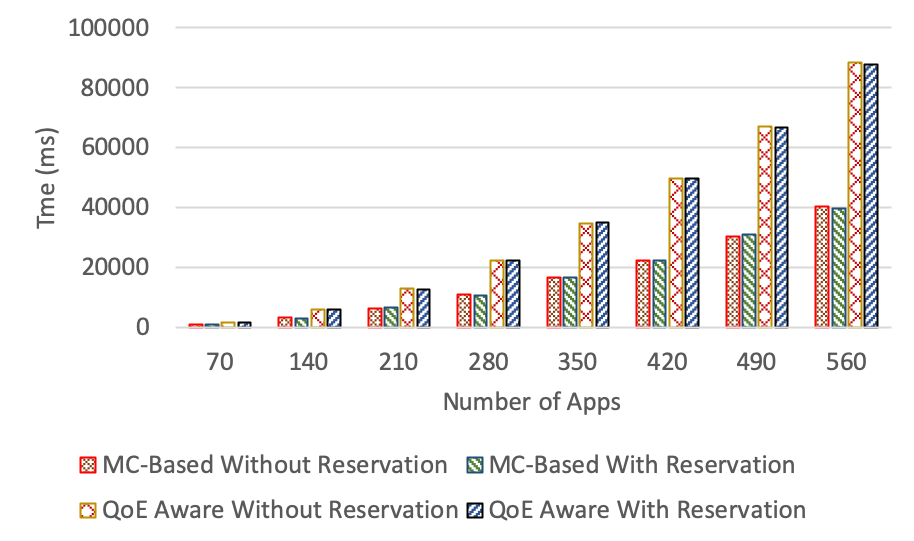}}
        \caption{Total delay}
        \label{2TotalDelay}
    \end{subfigure}
    \begin{subfigure}[b]{0.495\textwidth}
        \frame{\includegraphics[width=\textwidth]{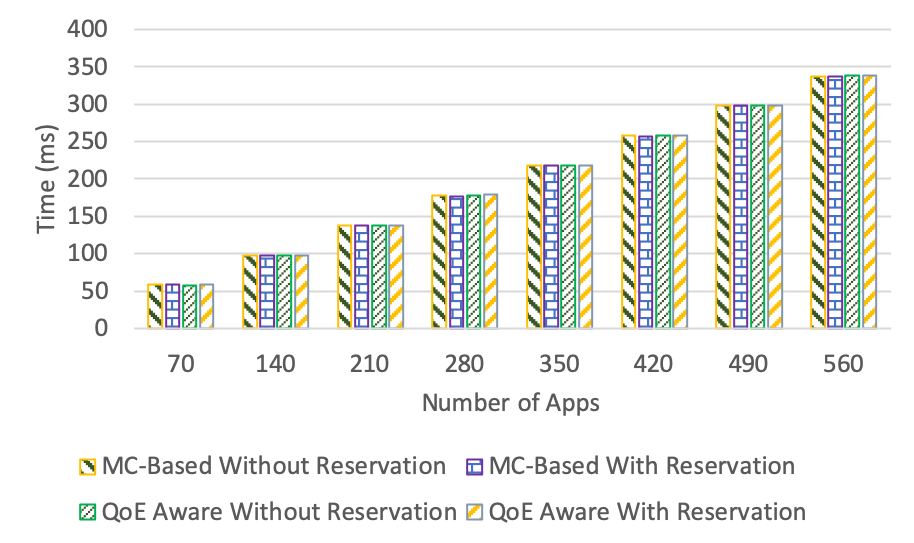}}
        \caption{Maximum delay}
        \label{3MaximumDelay}
    \end{subfigure}
    ~ 
    \begin{subfigure}[b]{0.495\textwidth}
        \frame{\includegraphics[width=\textwidth]{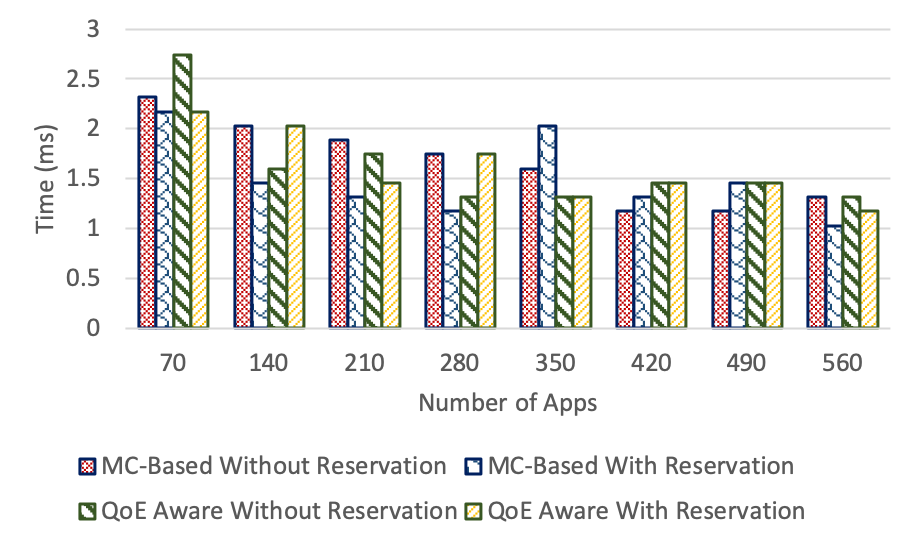}}
        \caption{Minimum delay}
        \label{4MinimumDelay}
    \end{subfigure}
    \caption{Average, total, maximum and minimum delays with the increasing number of user requests.}\label{fig:1Avg2Total3Max4Min}
\end{figure}

Based on the simulation results from the experiments, it was found that the average and the total delays of our proposed algorithm were less which is more important for time-sensitive applications. Fig. \ref{1AvgDelay} shows the average delay with the increasing number of task submissions. The average delay is decreased by 9\% on an average in the proposed MC-based policy when compared with QoE-aware policy. From the graph, it is clear that the average delay decreases linearly over the increasing number of task submissions. However, we cannot see a substantial difference between the with and without reservations in both algorithms. But resource reservations can affect SLA violation. In a scenario in which Fog device resources are busy with the tasks from other Fog clusters, the requests that come from the same Fog cluster will not be able to be served and this might cause SLA violations. 


We found approximately 51\% improvement in total delay by employing the proposed algorithm, as shown in Fig. \ref{2TotalDelay}. Since the Fog has been developed for time-sensitive applications, it is obvious that the improvement in total delay made the Fog environment usable for real implementation. However, maximum and minimum delay, processing time and cost should not be high. Therefore, we measured all of these parameters to evaluate the performance of our proposed algorithm.



Fig. \ref{3MaximumDelay} shows the maximum delay of both MC-based and QoE-aware policies. From the figure, it is clear that there is not much difference found in the maximum delay. There is approximately 0.04\% improvement in the proposed MC-based policy; this is negligible. 


The minimum delay of QoE-aware is better compared with the MC-based algorithm which is 4\% less on average, as illustrated in Fig. \ref{4MinimumDelay}. However, a minimum delay does not affect all applications; it is the minimum delay that is found after submitting all applications. Since, the total delay is more than half in the QoE-aware algorithm, the effect of a minimum delay can be overlooked.

\begin{figure}
    \centering
    \begin{subfigure}[b]{0.495\textwidth}
        \frame{\includegraphics[width=\textwidth]{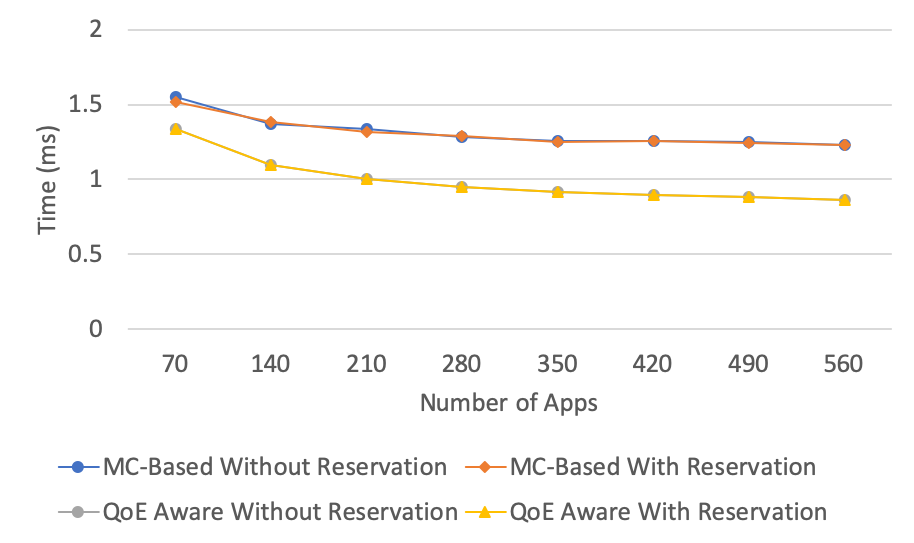}}
        \caption{Average processing time}
        \label{5AvgProcessingTime}
    \end{subfigure}
    ~ 
    \begin{subfigure}[b]{0.495\textwidth}
        \frame{\includegraphics[width=\textwidth]{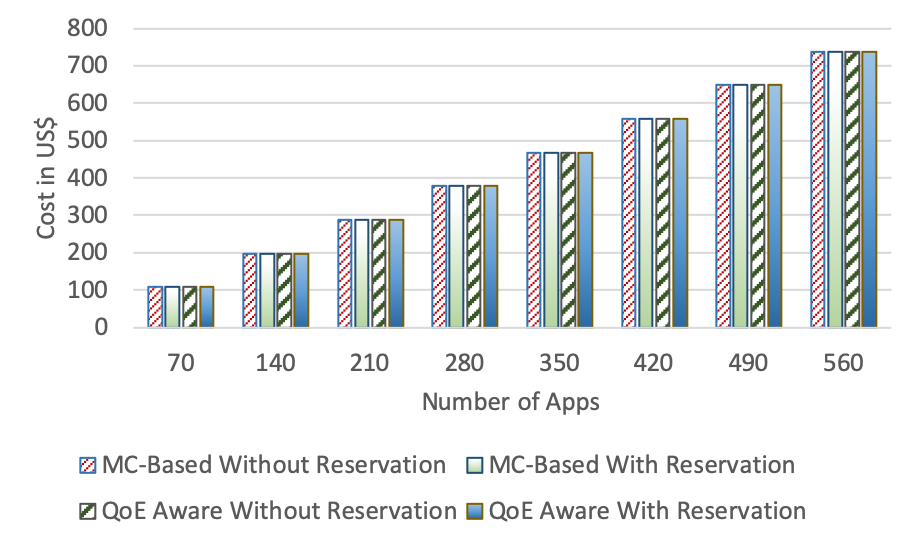}}
        \caption{Total processing costs}
        \label{6TotalProcessingCost}
    \end{subfigure}
    \caption{Average processing time and total processing costs with the increasing number of user requests.}\label{fig:5AvgPT6TotalPC}
\end{figure}


The average processing time of the QoE-aware algorithm is better because it is only focused on the processing time and speed, and round-trip time. It does not consider other Fog related assumptions, such as fluctuations in processing availability, distance and mobility of the Fog device. The average processing time of both algorithms is shown in Fig. \ref{5AvgProcessingTime}.


\begin{figure}[htbp]
	\centering
	\frame{\includegraphics[width=3.3in]{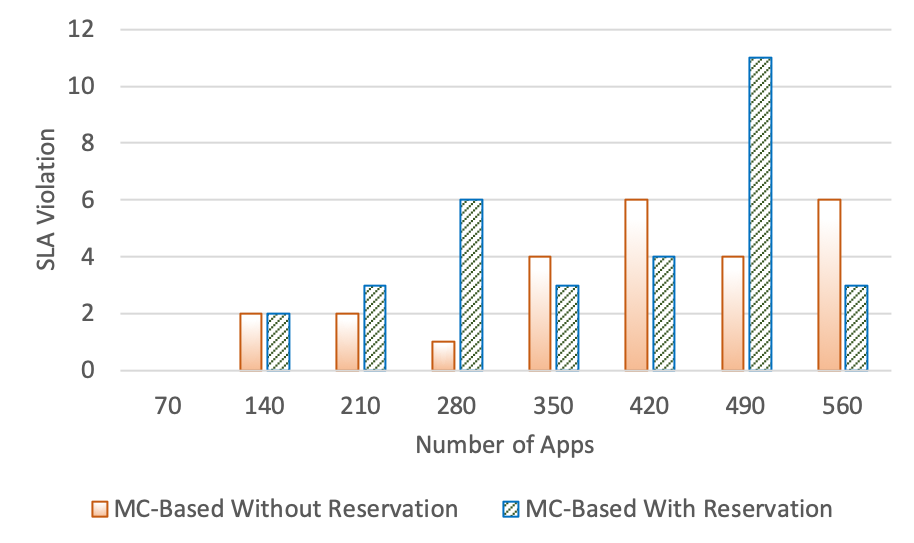}}
	\caption{Percentage of SLA violation with the increasing number of user request.}
	\label{7SLAVio}
\end{figure}

The cost of computation is shown in Fig. \ref{6TotalProcessingCost}. From the figure, it is clear that there is no difference in cost for both algorithms. We calculate the cost based on Amazon's IoT service cost which is based on messaging and connectivity cost, both of which are the same in simulation for both policies.

While the proposed algorithm is used for selecting resources for task execution without resource reservation, the SLA violation is high in most of the cases, as shown in Fig. \ref{7SLAVio}. This is because resources may be utilised to serve the other Fog devices which are subscribed to other peer Fog clusters. However, SLA violation decreases while resources have been reserved by employing a resource reservation algorithm. The penalty cost for SLA violation is 0.04\% without resource reservation and 0.03\% with resource reservation. 


\subsection{Independent Variation of Dynamic Behaviour Parameters}
In this experimental setting, we consider parameters in Table \ref{DybFogSimPara2} as default parameters, and we vary each parameter to examine the impact on the changes for each parameter, in order to understand how efficient the proposed policy is in a dynamic environment. We did not vary delay and distance during our simulations, since the delay always depends on the underlying network and device characteristics. Similarly, devices are scattered in a wireless network setting. It is not sensible to assume that multiple devices can be located at a similar distance.

\subsubsection{Impact on deadline-based user dynamic behaviour variation}
The percentage of changing deadline behaviour is examined to evaluate the performance of the proposed policy in a simulation environment. A range of 10\% to 80\% variation on the deadline was tested in a dynamic environment, with 10\% increments in variation each time. Lower variation means stricter deadlines, while higher variation means more flexible deadlines. All users vary the deadlines dynamically throughout the simulation. Other parameters such as free resources, battery availability, CPU utilisation fluctuation and distance varied dynamically within a range based on the default parameters. 

\begin{figure}[htbp]
    \centering
    \begin{subfigure}[b]{0.495\textwidth}
        \includegraphics[width=\textwidth]{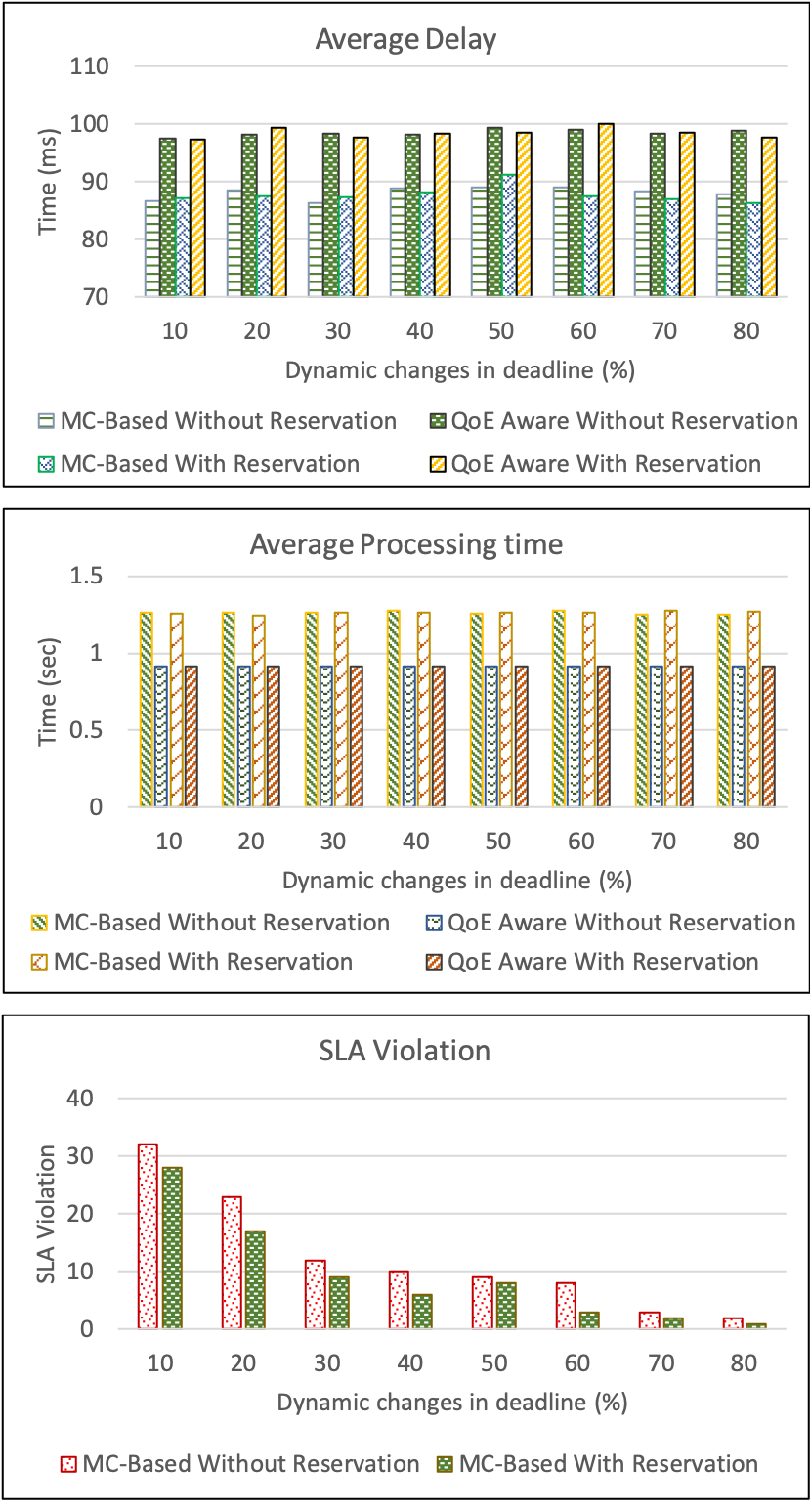}
        \caption{Dynamic deadline variation.}
        \label{8UserDynaDead}
    \end{subfigure}
    ~ 
    \begin{subfigure}[b]{0.495\textwidth}
        \includegraphics[width=\textwidth]{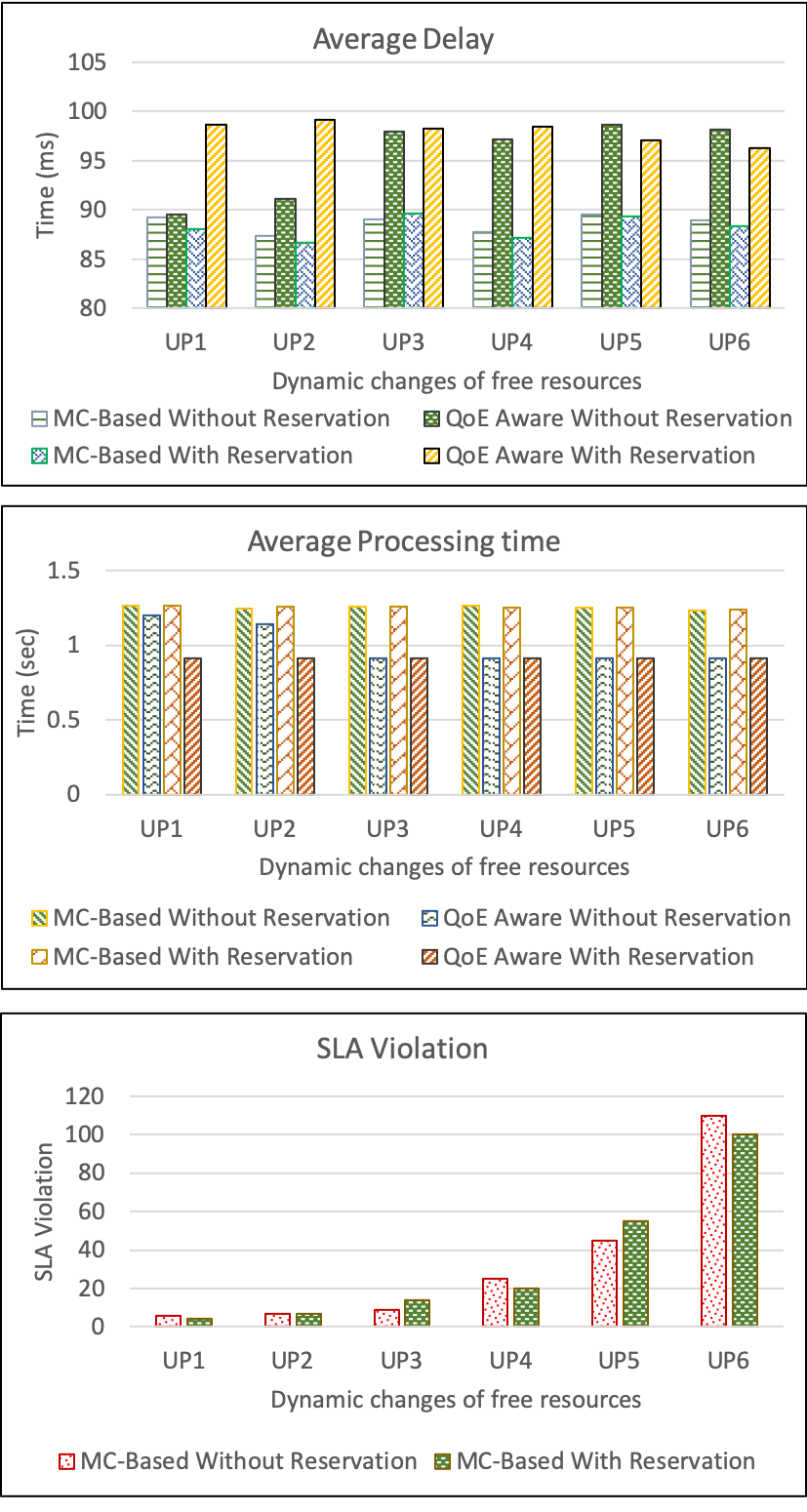}
        \caption{Dynamic free resource variation.}
        \label{9ResDynaFreeRes}
    \end{subfigure}
    \caption{Impact on delay, processing time and SLA violation in dynamic deadline variations and free resource variations.}\label{fig:8Udd9Rdf}
\end{figure}


Simulation results for user dynamic behaviour variation are shown in Fig. \ref{8UserDynaDead}. According to the figure, the proposed MC-Based policy has about 12\% less average delay compared with the QoE-aware policy. However, the average processing time for QoE-aware policy is lower. But again, the QoE-aware policy did not consider SLA violation in their work. SLA violation in without reservation is about 60\% higher on average, compared with policy with reservation. This is because, Fog devices might be busy with other requests from the peer Fog device, so they are unable to serve their own application requests.    

\subsubsection{Impact on free resource variation}
To examine free resource variation, variation ranges take place from 0\% to 60\%. While variation is over 60\%, the SLA violation is very high. Due to that, we should not register that the Fog device with free resource variation is more than 60\%.  Variation is categorised as UP1, UP2, UP3, UP4, UP5 and UP6, which represents variations between 0\% to 10\%, 10\% to 20\%, 20\% to 30\%, 30\% to 40\%, 40\% to 50\% and 50\% to 60\%, respectively. Other parameters, such as deadline, battery availability, CPU utilisation fluctuation and distance, varied dynamically within a range based on the default parameters.


\begin{figure}[htbp]
    \centering
    \begin{subfigure}[b]{0.495\textwidth}
        \includegraphics[width=\textwidth]{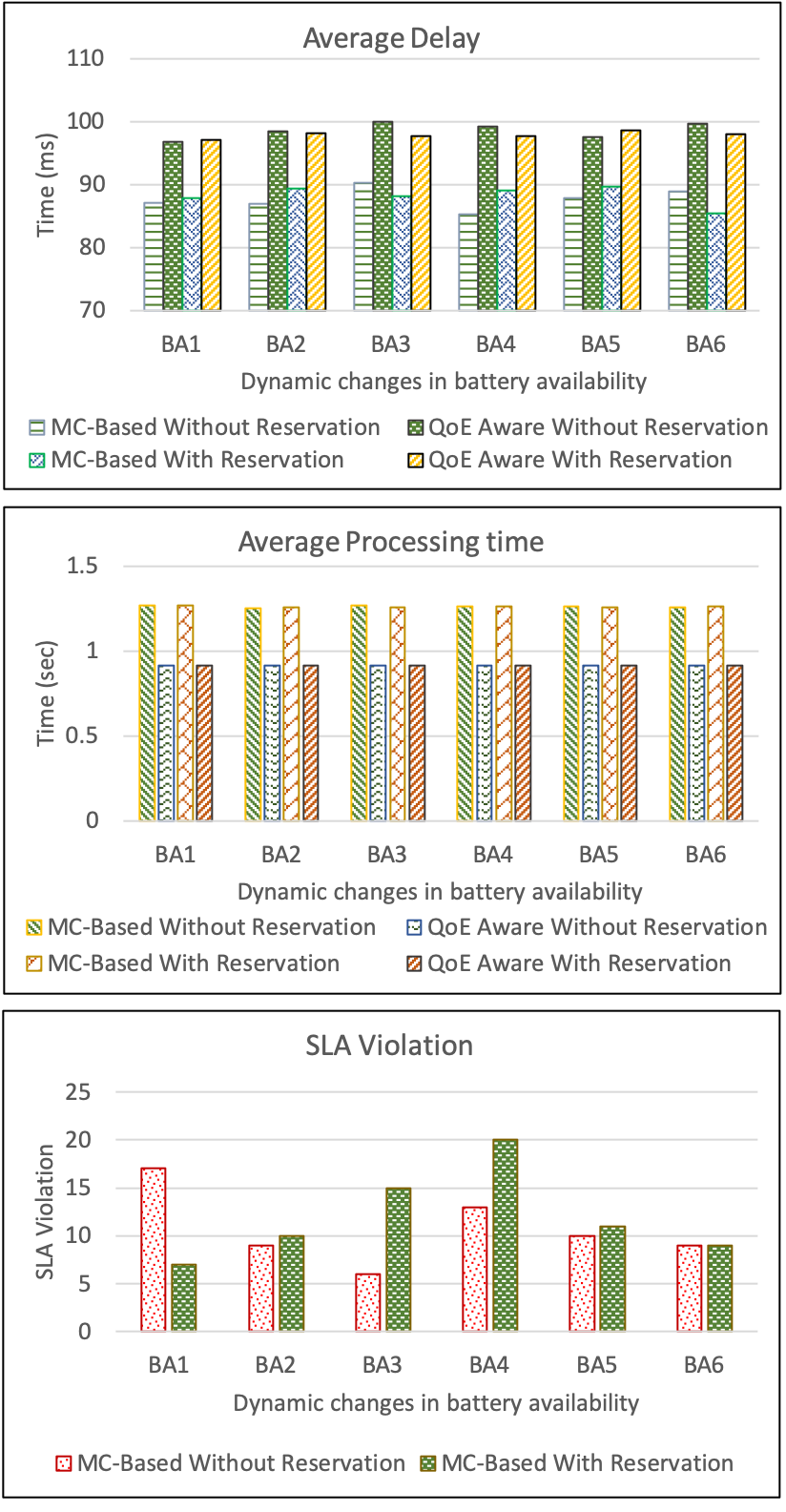}
        \caption{Dynamic battery power variation.}
        \label{10ResDynaBatAval}
    \end{subfigure}
    ~ 
    \begin{subfigure}[b]{0.495\textwidth}
        \includegraphics[width=\textwidth]{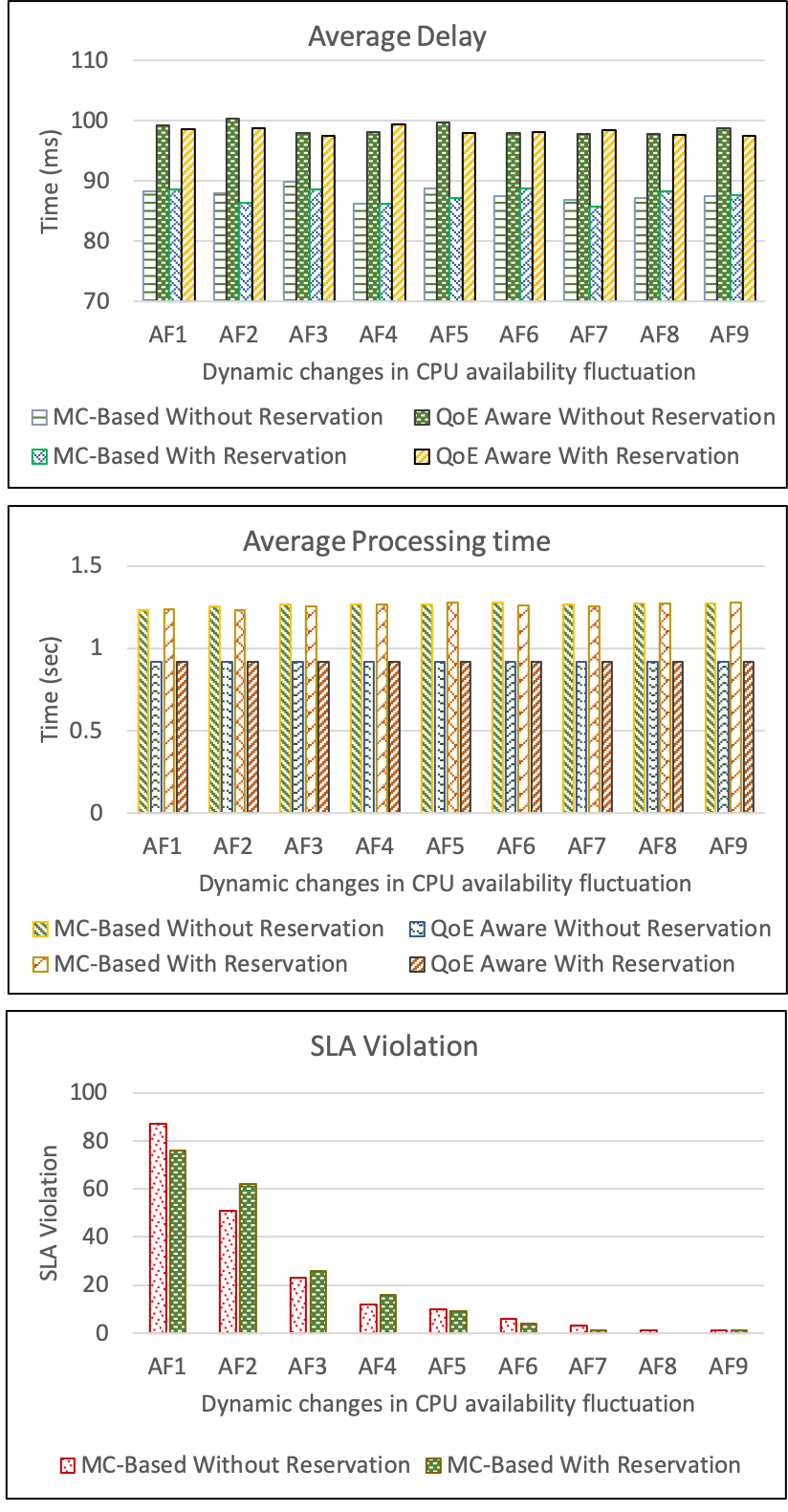}
        \caption{Dynamic CPU utilisation fluctuation.}
        \label{11ResDynaCPUAvalFl}
    \end{subfigure}
    \caption{Impact on delay, processing time and SLA violation in dynamic battery power variation and CPU utilisation fluctuation.}
    \label{fig:10Rdba11Rdca}
\end{figure}

Simulation results for free resources are shown in Fig. \ref{9ResDynaFreeRes}. In the figure, the proposed MC-Based policy has around 8\% and 11\% less average delay for without and with reservation respectively, compared with QoE-aware policy. However, average processing time for QoE-aware policy is lower which did not consider SLA violation. SLA violation in without reservation is about 5\% higher on an average, compared with policy with reservation, while free resource is changes dynamically.

\subsubsection{Impact on available battery power variation}
Fog processing will have a negative impact if battery availability of the Fog device is low. To evaluate the impact on battery power, we varied battery power availability from 0\% to 90\%, by varying 15\% available battery power each time. The variations are represented as BA1, BA2, BA3, BA4, BA5 and BA6. Other parameters such as deadlines, free resources, CPU utilisation fluctuation and distance varied dynamically within a range, based on the default parameters.


Simulation results for available battery power variations are shown in Fig. \ref{10ResDynaBatAval}. As shown in the figure, the proposed MC-Based policy has around 12\% and 11\% less average delay for without and with reservation, respectively, compared with the QoE-aware policy. Although the average processing time for QoE-aware policy is lower, the work did not consider SLA violation. SLA violation in without reservation is about 5\% higher on an average, compared with policy with reservation, while available battery power varies dynamically. 

\subsubsection{Impact on CPU utilisation fluctuation variation}
Fluctuation on CPU availability in Fog devices is examined by varying CPU availability fluctuation. The variation started from 0\% to 50\%, in increments of 10\% each time without changing the lower boundary, which is represented as AF1 to AF9. Other parameters such as deadlines, free resources, battery availability and distance varied dynamically within a range based on the default parameters.


Simulation results for CPU utilisation fluctuation are shown in Fig. \ref{11ResDynaCPUAvalFl}. According to the figure, the proposed MC-Based policy has more than 12\% less average delay compared with the QoE-aware policy for both with and without reservation policy. Average processing time for MC-Based policy is higher compared with the QoE-aware policy, which is similar to other variation results. SLA violation in without reservation is over 24\% higher, on average, compared with policy with reservation, while CPU utilisation fluctuates dynamically.

From all the above results, we can conclude that the proposed MC-based policy performs better compared with the QoE-aware policy, in most cases. Consideration of SLA violation is important in the Fog because it has evolved to serve time-sensitive applications.

\section{Conclusion}
Placing applications in the Fog environment is challenging because of the dynamic nature of users and computation devices. In the Fog, we cannot be sure that the application which has been submitted to a device can be completed by the same device. On the other hand, user requirements might change in the meantime. Hence, we need to consider a number of Fog computing characteristics before placing an application in the Fog environment. We propose an MC-based resource allocation algorithm in the Fog environment which will take care of the characteristics of Fog computing, as well as user dynamic behaviour after application submission. The proposed algorithm is further refined by applying resource reservation, in order to minimise SLA violation. Experimental results show that the proposed algorithm is performing better than the existing one. On average, the total delay of the proposed algorithm was reduced by 51\% compared with the QoE-aware algorithm which can better serve for time-sensitive applications. For the future work, we will employ our proposed method in a real Fog computing environment, instead of simulation. In this work we did not consider failure of the Fog devices which we will consider in the future.

\begin{acks}
The first author would like to thank University of Tasmania (UTAS) for providing Tasmania Graduate Research Scholarship (TGRS) for supporting his studies.
\end{acks}

%
\bibliographystyle{ACM-Reference-Format}
\bibliography{sample-base}

%
\appendix









\end{document}